\documentclass[11pt,a4paper]{article}
\usepackage{jheppub}
\usepackage{amssymb}   % for math
\usepackage{amsmath}
\usepackage{graphicx}
\usepackage{caption}
\usepackage{subcaption}
\usepackage{axodraw4j}
\usepackage{mathrsfs}
\usepackage{url}

\def\beq{\begin{equation}}   
\def\eeq{\end{equation}}
\def\bea{\begin{eqnarray}}  
\def\eea{\end{eqnarray}}

\def\f21{{}_2F_{1}}
\def\beq{\begin{equation}}
\def\eeq{\end{equation}}
\def\bsp#1\esp{\begin{split}#1\end{split}}

\newcommand{\cA}{{\mathcal{A}}}
\newcommand{\ord}{{\mathcal{O}}}

\catcode`\@=11
\font\manfnt=manfnt
\def\Watchout{\@ifnextchar [{\W@tchout}{\W@tchout[1]}}
\def\W@tchout[#1]{{\manfnt\@tempcnta#1\relax%
  \@whilenum\@tempcnta>\z@\do{%
    \char"7F\hskip 0.3em\advance\@tempcnta\m@ne}}}
\let\foo\W@tchout
\def\dubious{\@ifnextchar[{\@dubious}{\@dubious[1]}}

\def\@dubious[#1]{%
  \setbox\@tempboxa\hbox{\@W@tchout#1}
  \@tempdima\wd\@tempboxa
  \list{}{\leftmargin\@tempdima}\item[\hbox to 0pt{\hss\@W@tchout#1}]}
\def\@W@tchout#1{\W@tchout[#1]}
\catcode`\@=12
\title{Higgs production in bottom quark fusion: Matching the 4- and 5-flavour schemes to third order in the strong coupling}

\author[a]{Claude Duhr}
\author[b]{Falko Dulat}
\author[c]{Valentin Hirschi}
\author[b,d]{Bernhard Mistlberger}
\affiliation[a]{Theoretical Physics Department, CERN, CH-1211 Geneva 23, Switzerland.}
\affiliation[b]{SLAC National Accelerator Laboratory, Stanford University, Stanford, CA 94039, USA.}
\affiliation[c]{ETH Z\"urich, R\"amistrasse 101, 8092 Z\"urich, Switzerland.}
\affiliation[d]{Center for Theoretical Physics, Massachusetts Institute of Technology, Cambridge, MA 02139, USA.}

\emailAdd{claude.duhr@cern.ch, falko.dulat@gmail.com, hirschva@itp.phys.ethz.ch, bernhard.mistlberger@gmail.com}

\preprint{CERN-TH-2020-051,MIT-CTP/5191}
\abstract{
We present analytic results for the partonic cross sections for the production of a Higgs boson via the fusion of two bottom quarks at N$^3$LO in QCD perturbation theory in the five-flavour scheme.
We combine this perturbative result with NLO accurate predictions in the four-flavour scheme that include the full bottom quark mass dependence by appropriately removing any double-counting stemming from contributions included in both predictions.
We thereby obtain state-of-the-art predictions for the inclusive production probability of a Higgs boson via bottom quark fusion at hadron colliders.
}


\begin{document}

\maketitle

% !TEX root = paper.tex

\section{Introduction}
\label{sec:intro}
Measuring precisely the properties of the Higgs boson, and possibly establishing the Standard Model (SM) of particle physics as the correct mechanism to explain the electroweak symmetry breaking, is one of the primary goals of the third run of the Large Hadron Collider (LHC) and its future upgrades. Since the SM Higgs boson couples to other particle species with a coupling strength proportional to their mass, measurements of the couplings of the Higgs boson to massive electroweak bosons and third generation fermions - the $\tau$ lepton as well as top and bottom quarks -- are promising candidates to probe its interactions. The Yukawa coupling of the bottom quark is of particular interest, as several models of New Physics -- like for example minimal supersymmetric extensions of the Standard Model -- predict enhanced bottom Yukawa couplings (see chapter IV.2.2 of ref.~\cite{deFlorian:2016spz}).

The interactions of the Higgs boson and the bottom quark can be probed at the LHC either through processes in which the Higgs decays to a pair of bottom quarks, or through processes in which it is produced from bottom quarks. 
In principle it is possible to directly constrain the bottom quark Yukawa coupling by measuring the decay of a Higgs boson into a bottom quark pair. However, even though this decay benefits from a large branching fraction, it is challenging to measure it precisely at a hadron collider due to the purely hadronic final state signature~\cite{Aaboud:2018zhk,Sirunyan:2018kst}. Moreover, any measurement of a Higgs boson decay necessarily relies on a precise prediction for its inclusive production cross-section. It is thus beneficial to study Higgs production processes at the LHC that involve bottom quarks. To measure the Yukawa coupling in this fashion, one particularly relevant production mechanism is that of the annihilation of two bottom quarks extracted from the colliding hadrons.
The goal of this paper is thus to perform a phenomenological study of the production of a Higgs boson through bottom quark fusion.

Due to the small, but non-negligible, value of the bottom quark mass, there are two different ways in which one can model theoretical predictions for LHC processes involving bottom quarks. In the five-flavour scheme, the bottom quark is considered a massless parton. Consequently, all finite-mass effects are neglected, except for collinear logarithms that are resummed into the parton density functions. The five-flavour scheme has the advantage that the computation of higher-order corrections in the strong coupling constant is greatly simplified because all relevant quark species are massless (we neglect all top quark effects in the computations performed in the five-flavour scheme). In this scheme the inclusive bottom quark fusion cross section was computed through next-to-next-to-leading order (NNLO) already almost two decades ago~\cite{Dicus:1998hs,Balazs:1998sb,Harlander:2003ai}. Very recently, a subset of the authors have computed for the first time the next-to-next-to-next-to-leading order (N$^3$LO) corrections~\cite{Duhr:2019kwi} (for a combination of the N$^3$LO cross section with resummation of threshold logarithms, see ref.~\cite{H:2019dcl}). In a first part of this paper, we give more details on the structure of the partonic coefficient functions of ref.~\cite{Duhr:2019kwi}. In particular, we make all the partonic coefficient functions publicly available as ancillary material attached to the arXiv submission of this paper. We also perform a detailed phenomenological analysis of Higgs production in bottom quark fusion, and we investigate the main sources of uncertainty that affect the cross section at N$^3$LO. 

While effects due to the non-zero mass of the bottom quark are expected to be small, they can nevertheless lead to sizeable effects, especially when compared to the level of precision with which the QCD effects are incorporated at N$^3$LO. In the four-flavour scheme the bottom quark is treated as massive and is produced in the hard process, leading to higher final-state multiplicities. Consequently, Higgs production in bottom quark fusion is only known through next-to-leading order (NLO) in the four-flavour scheme~\cite{Dittmaier:2003ej,Dawson:2003kb,Wiesemann:2014ioa}. As massive quarks cannot appear as initial state partons, all bottom quarks are generated from gluon splittings. While the non-zero mass protects the gluon splittings from collinear divergences, the four-flavour scheme is plagued by large logarithms involving the bottom quark mass which may spoil the convergence of the perturbative series. It is therefore desirable to combine the two schemes into a single prediction. Several methods to perform this combination have been proposed in the literature, ranging from purely phenomenological prescriptions~\cite{Harlander:2011aa} to theoretically well-grounded matching procedures~\cite{Bonvini:2015pxa,Bonvini:2016fgf,Forte:2015hba,Forte:2016sja}. 
So far, however, all these prescriptions have suffered from the fact that the equivalent of the NNLO result in the five-flavour scheme is only the leading order cross section in the four-flavour scheme. 
No matched prediction including all ingredients consistently through third order in the strong coupling has been obtained. 

One of the main results of our paper is the first consistent matching of the four and five-flavour schemes through third order in the strong coupling. 
This is made possible by combining the N$^3$LO result for the cross section of ref.~\cite{Duhr:2019kwi} with the matching procedure of refs.~\cite{Forte:2015hba,Forte:2016sja}. 
In this way we are able to obtain the most precise predictions for this process, where all QCD and mass effects are included through third power in the strong coupling, and all logarithms of the bottom quark mass are resummed at leading power through next-to-next-to-leading logarithmic (NNLL) accuracy.

Our paper is organised as follows: In section~\ref{sec:setup} we review inclusive Higgs production in the four and five-flavour schemes, and we introduce our notations and conventions. In section~\ref{eq:coeff_funcs} we discuss the analytic structure of the partonic coefficient functions at N$^3$LO in the five-flavour scheme, and in section~\ref{sec:5flav} we present a detailed analysis of the different sources of uncertainty that affect the N$^3$LO cross section. In section~\ref{sec:fonnll} we review the FONLL matching scheme, and in section~\ref{sec:matched_pheno} we present our results for the combination of the two schemes. In section~\ref{sec:conclusion} we draw our conclusions.

% !TEX root = paper.tex

\section{Setup of the computation}
\label{sec:setup}

\subsection{Higgs production in bottom quark fusion}

In this section we review some basic facts about Higgs production in bottom quark fusion, and we introduce our notations and conventions. Using QCD factorisation, the cross section can be written as
\beq\label{eq:qcd_fac}
\sigma_{P\, P\rightarrow H+X} = \int_0^1 dx_1\,dx_2\,\sum_{i,j} f_i(x_1,\mu_F^2)f_j(x_2,\mu_F^2)\,\hat{\sigma}_{ij}\,,
\eeq
where $\mu_F$ denotes the factorisation scale and the $f_i(x,\mu_F^2)$ denote the parton density functions (PDFs) to find a parton species $i$ with momentum fraction $x$ inside the proton. The $\hat{\sigma}_{ij}$ denote the partonic cross sections to produce a Higgs boson from a collision of two partons $i$ and $j$. Here we are interested in the production of a Higgs boson from the fusion of a pair of bottom quarks. More precisely, we focus on the part of the cross section proportional to $y_b^2$, where $y_b$ denotes the bottom quark Yukawa coupling. The sum runs over all active partons in the proton, i.e. gluons and all massless quark flavours. 

Due to the small mass $m_b$ of the $b$ quark compared to the mass $m_H$ of the Higgs boson, there are two ways in which eq.~\eqref{eq:qcd_fac} can be interpreted. In the four-flavour scheme (4FS) the bottom quark is considered massive. Consequently, there is no PDF for the bottom quark and all finite mass effects are retained in the partonic cross sections. The non-zero mass also prevents the appearance of collinear singularities involving $b$ quarks. Instead, the partonic cross sections develop collinear logarithms $\log Q^2/m_b^2$, where $Q\sim m_H$ denotes the hard scale of the process. Given the hierarchy between the Higgs and the bottom quark masses, these logarithms may spoil the convergence of the perturbative series and need to be resummed to all orders in perturbation theory.
This resummation is achieved by working in the five-flavour scheme (5FS), where the bottom quark is treated as massless and interpreted as a parton inside the proton. While the 5FS has the advantage that all collinear logarithms are resummed into the bottom quark PDF, it suffers from the fact that, unlike in the 4FS, the cross sections in the 5FS do not include any finite-$m_b$ non-logarithmic effects.
\begin{center}
\begin{table}[!t]
\begin{center}
\begin{tabular}{c|c|c|c|c}
\hline\hline
 & \includegraphics{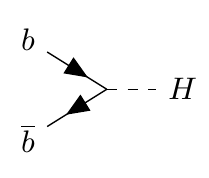} & \includegraphics{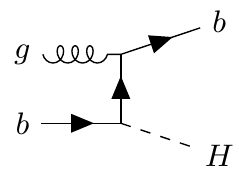} & \includegraphics{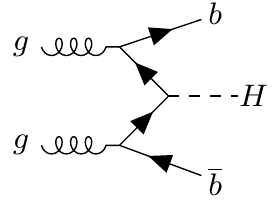} & \includegraphics{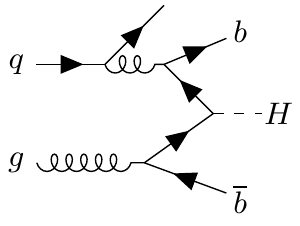}\\
 \hline &&&&\\
  $\quad$4FS$\quad$ & -- & -- & LO & NLO \\
 $\quad$5FS$\quad$ & LO & NLO & NNLO & N$^3$LO
 % \\
 %Partonic channels (5FS) & $b\bar{b}$ & $b\bar{b},bg$ & $b\bar{b},bg,bb,bq,b\bar{q},gg,q\bar{q}$ & $b\bar{b},bg,bb,bq,b\bar{q},gg,q\bar{q},qg$\\
 %&&&&
 \\
  \hline\hline
\end{tabular}
\caption{\label{fig:diagrams}  Representative diagrams contributing at different orders in perturbation theory in the 4FS and 5FS. 
%The last line summarises the partonic channels in the 5FS. Channels related by charge conjugation are not shown explicitly and $q$ denotes a light quark that does not couple directly to the Higgs boson. The partonic channels in the 4FS are obtained by ignoring initial states involving a bottom quark.
}
 \end{center}
 \end{table}
 \end{center}

The 4FS and 5FS start to contribute at different orders in the perturbative expansion in the strong coupling constant $\alpha_s$. Indeed, in the 4FS (and under the assumption that there is no intrinsic bottom quark in the proton) the bottom quarks are generated perturbatively from gluon splittings, and therefore the perturbative expansion in the 4FS starts at order $\alpha_s^2$. In the 5FS, instead, the bottom quark is considered a parton, and the leading-order cross section is proportional to $\alpha_s^0$. Representative Feynman diagrams that contribute to each of the two schemes are shown in tab.~\ref{fig:diagrams}.

The inclusive cross section in the 4FS can be written as
\beq\label{eq:sigma_4FS}
\sigma^{(4)} = \tau\hat \sigma_0(y_b^2,m_H^2) \sum_{i,j=-4}^4\,\mathscr{L}_{ij}^{(4)}(\tau,\mu_F^2)\otimes {\eta}_{ij}^{(4)}\!\left(\tau,L_f,L_r,m_b,\alpha_s^{(4)}\right)\,.
\eeq
Here $\tau = \frac{m_H^2}{S}$, with $S$ the hadronic center-of-mass energy, and $y_b\equiv y_b(\mu_R^2)$ and $\alpha_s^{(4)}\equiv \alpha_s^{(4)}(\mu_R^2)$ denote the Yukawa coupling of the $b$-quark and the strong coupling constant for $N_f=4$ massless quark flavours.
In the 4FS computation, the strong coupling is renormalised in the mixed scheme of ref.~\cite{Collins:1978wz} in which the contribution from the four massless quark flavours is subtracted in the $\overline{\text{MS}}$ scheme, while the contribution from the massive bottom and top quarks running in the fermionic loop of the one-loop gluon self-energy is subtracted on-shell.
We define the normalisation factor
\beq
\label{eq:sigma0}
\hat{\sigma}_0(y_b^2,m_H^2) =\frac{ m_b^2 \pi }{2n_c v^2 m_H^2},\hspace{1cm} m_b^2=y_b^2 v^2.
\eeq
Here, $v$ is the vacuum expectation value of the Higgs field and $n_c$ refers to the number of colours.
The renormalisation and factorisation scales are denoted by $\mu_R$ and $\mu_F$ respectively. 
Unless specified otherwise, all coupling constants are evaluated at a renormalisation scale $\mu_R$. 
The partonic luminosities are defined as the convolution of the corresponding four-flavour PDFs,
\beq
\mathscr{L}_{ij}^{(4)}(\tau,\mu_F^2) = f^{(4)}_i(\tau,\mu_F^2)\otimes  f^{(4)}_j(\tau,\mu_F^2)\,,
\eeq
where the convolution is defined by
\beq
f(x)\otimes g(x) = \int_x^1\frac{dz}{z}\,f(z)\,g\left(\frac{x}z{}\right)\,.
\eeq
The sum in eq.~\eqref{eq:sigma_4FS} runs over all four massless quark flavours and the gluon. We find it convenient to use both integer numbers and explicit parton names as indices, e.g.:
\begin{equation}
\label{eq:labeling_conventions}
\{f_{-5},f_{-4},f_{-3},f_{-2},f_{-1},f_0,f_1,f_2,f_3,f_4,f_5\} \equiv \{f_{\bar{b}},f_{\bar{c}},f_{\bar{s}},f_{\bar{u}},f_{\bar{d}},f_g,f_d,f_u,f_s,f_c,f_b\}.
\end{equation}
The partonic coefficient functions depend on the bottom quark pole mass and the logarithms of the factorisation and renormalisation scales:
\beq
L_f = \log\frac{\mu_F^2}{m_H^2} \textrm{~~~and~~~}L_r = \log\frac{\mu_F^2}{\mu_R^2}\,.
\eeq
They admit the perturbative expansion:
\beq\bsp
{\eta}_{ij}^{(4)}\!&\left(z,L_f,L_r,m_b,\alpha_s^{(4)}\right) = \sum_{n=2}^{\infty}a^{(4)n}_s\,{\eta}_{ij}^{(4,n)}\!\left(z,L_f,L_r,m_b\right)\,,
\esp\eeq
with $a^{(4)}_s\equiv a^{(4)}_s(\mu_R^2) = \alpha^{(4)}_s(\mu_R^2)/\pi$. The partonic coefficient functions in the 4FS are known (numerically) through NLO~\cite{Dittmaier:2003ej,Dawson:2003kb,Wiesemann:2014ioa}.

Similarly, the inclusive cross section for Higgs production in bottom quark fusion in the 5FS can be cast in the form
\beq\label{eq:sigma_5FS}
\sigma^{(5)} = \tau\hat{\sigma}_0(y_b^2,m_H^2) \sum_{i,j=-5}^5\,\mathscr{L}_{ij}^{(5)}(\tau,\mu_F^2)\otimes {\eta}_{ij}^{(5)}\left(\tau,L_f,L_r,\alpha_s^{(5)}\right)\,.
\eeq
Above, we again chose to normalise the partonic coefficient functions in the 5FS by the factor $\hat{\sigma}_0$ defined in eq.~\eqref{eq:sigma0}.
Throughout this paper we use the convention that $X^{(n)}$ denotes the quantity $X$ computed in the $n$ flavour scheme, and the notations introduced for the 4FS remains valid in the 5FS context. The main difference between the cross sections in the 4FS and 5FS in eqs.~\eqref{eq:sigma_4FS} and~\eqref{eq:sigma_5FS} is that in the 4FS the partonic coefficient functions have an explicit dependence on the bottom (pole) mass $m_b$, and that the 4FS expression does not include the bottom quark into the sum over flavours. In particular, the coefficient functions $\eta_{ij}^{(5)}$ admit the perturbative expansion
\beq\label{eq:eta_5FS}
\eta_{ij}^{(5)}\left(z,L_f,L_r,\alpha_s^{(5)}\right)=\sum\limits_{n=0}^\infty a^{(5)n}_s\,\eta_{ij}^{(5,n)}\left(z,L_f,L_r\right).
\eeq
The partonic coefficient functions in the 5FS are known at NLO~\cite{Dicus:1998hs,Balazs:1998sb} and NNLO~\cite{Harlander:2003ai}. Very recently also the N$^3$LO corrections have become available~\cite{Duhr:2019kwi}. We will review the results of ref.~\cite{Duhr:2019kwi} in the next section.

\section{Partonic coefficient functions in the 5FS}
\label{eq:coeff_funcs}
One of the main results of this paper are expressions for the N$^3$LO corrections to the partonic coefficient functions $\eta_{ij}^{(5)}$ for the production of a Higgs boson in bottom quark fusion. 
In this section we first discuss the general structure and computation of the partonic cross sections.
We then explain the function space needed to represent the partonic coefficient functions.
Finally, we give an alternative representation of our partonic coefficient functions in terms of expansions around different expansion points.

\subsection{Structure of the partonic coefficient functions}

%The partonic coefficient function can be expanded in the strong coupling constant as 
%\VHcomment{Use $a_s$ instead of $\frac{\alpha_s}{\pi}$ since it has already been introduced?}
%\beq
%\eta_{ij}^{(5)}\left(z,L_f,L_r,\alpha_s^{(5)}\right)=\sum\limits_{n=0}^\infty\left(\frac{\alpha_s^{(5)}(\mu_R)}{\pi}\right)^n\eta_{ij}^{(5,n)}\left(z,L_f,L_r\right).
%\eeq
At LO the only non-vanishing partonic coefficient functions have a bottom and anti-bottom quark in the initial state:
\beq
\eta_{b\bar b}^{(5,0)}\left(z,L_f,L_r\right)=\eta_{\bar bb}^{(5,0)}\left(z,L_f,L_r\right)=\delta(1-z)\,.
\eeq
The variable $z$ is defined by
\beq
z=\frac{m_H^2}{x_1 x_2 S}\,,
\eeq
where the $x_i$ are defined in eq.~\eqref{eq:qcd_fac}.
Up to third order in the strong coupling constant there are eight distinct functions necessary in order to describe all partonic coefficient functions for different initial states.
These eight functions are given by
\beq
\left\{
\eta_{b\bar b}^{(5,n)},\hspace{0.3cm}
\eta_{bg}^{(5,n)},\hspace{0.3cm}
\eta_{bq}^{(5,n)},\hspace{0.3cm}
\eta_{b\bar q}^{(5,n)},\hspace{0.3cm}
\eta_{bb}^{(5,n)},\hspace{0.3cm}
\eta_{gg}^{(5,n)},\hspace{0.3cm}
\eta_{q\bar q}^{(5,n)},\hspace{0.3cm}
\eta_{qg}^{(5,n)}
\right\}.
\eeq
Above, $g$, $\bar b$ and $b$ refer to a gluon, anti-bottom quark and bottom quark respectively, and $q$ and $\bar q$ refer to a single quark and anti-quark that is not a bottom (anti-) quark.
Results for the partonic coefficient functions at NLO and NNLO were computed in refs.~\cite{Dicus:1998hs,Balazs:1998sb,Harlander:2003ai}.
The above functions were obtained by a subset of the authors at N$^3$LO for the purposes of ref.~\cite{Duhr:2019kwi}.
Here, we present explicit results for these functions and make them publicly available in computer-readable form as ancillary material of this article.

The computation of the the N$^3$LO partonic coefficient functions follows the same strategy as that of the computation of the inclusive cross section for Higgs boson production through gluon fusion~\cite{Anastasiou:2015ema,Mistlberger:2018etf} and the inclusive Drell-Yan cross section~\cite{Duhr:2020seh}.
In particular, the results were obtained by using the framework of reverse unitarity~\cite{Anastasiou2002,Anastasiou2003,Anastasiou:2002qz,Anastasiou:2003yy,Anastasiou2004a} in order to compute all required interferences of real and virtual amplitudes contributing to the N$^3$LO cross section. 
The required phase-space and loop integrals were carried out implicitly by using integration-by-part (IBP) identities~\cite{Tkachov1981,Chetyrkin1981,Laporta:2001dd}  together with the method of differential equations~\cite{Kotikov:1990kg,Kotikov:1991hm,Kotikov:1991pm,Henn:2013pwa,Gehrmann:1999as}. 
This method allows one to represent the required integrated and interfered amplitudes in terms of linear combinations of \emph{master integrals}.
Purely virtual amplitudes were first computed in ref.~\cite{Gehrmann:2014vha} using the master integrals from refs.~\cite{Gehrmann:2006wg,Heinrich:2007at,Heinrich:2009be,Lee:2010cga,Baikov:2009bg,Gehrmann:2010ue,Gehrmann:2010tu}, and  recomputed and confirmed in ref.~\cite{Duhr:2019kwi}.
Contributions with one real parton in the final state were considered in refs.~\cite{Anastasiou:2013mca,Kilgore:2013gba,Duhr:2013msa,Li:2013lsa,Dulat:2014mda,Ahmed:2014pka} and the master integrals we used for our calculation were documented in refs.~\cite{Dulat:2014mda,Anastasiou:2013mca}.
Master integrals with two and three real partons were obtained for the purpose of ref.~\cite{Mistlberger:2018etf} and are based on results from refs.~\cite{Anastasiou:2014vaa,Li:2014bfa,Li:2014afw,Anastasiou:2015yha,Anastasiou:2013srw,Anastasiou:2015ema}.

We work in the $\overline{\text{MS}}$-scheme in conventional dimensional regularisation. 
The ultraviolet (UV) counterterm for the strong coupling constant has been determined through five loops in refs.~\cite{Tarasov:1980au,Larin:1993tp,vanRitbergen:1997va,Baikov:2016tgj,Herzog:2017ohr}.
The renormalisation constant for the Yukawa coupling is identical to the quark mass renormalisation constant of QCD in the $\overline{\textrm{MS}}$-scheme~\cite{Harlander:2003ai,vanRitbergen:1997va,Chetyrkin:1997dh,Czakon:2004bu,Baikov:2014qja}. 
Infrared (IR) divergences are absorbed into the definition of the PDFs using mass factorisation at N$^3$LO~\cite{Buehler:2013fha,Hoschele:2012xc,Hoeschele:2013gga}.
The mass factorisation involves convoluting lower-order partonic cross sections with the three-loop splitting functions of refs.~\cite{Moch:2004pa,Vogt:2004mw,Ablinger:2017tan}. 
We have computed all the convolutions analytically in $z$ space using the {\sc PolyLogTools} package~\cite{Duhr:2019tlz}.
After combining our interfered matrix elements with the UV and PDF-IR counterterms we send the dimensional regulator to zero and obtain our final results.

The partonic coefficient functions for a bottom and anti-bottom quark in the initial state contain distributions in the variable $z$ that were already obtained in ref.~\cite{Ahmed:2014cha}. We checked that our computation agrees with this result.
We refer to these contributions as soft-virtual (SV) contributions and to the non-distribution-valued part of the partonic coefficient functions as the regular part. 
Consequently, we split our partonic coefficient functions into regular and SV parts.
\beq
\eta_{ij}^{(5,n)}\left(z,L_f,L_r\right)=\eta_{ij,\,\text{SV}}^{(5,n)}\left(z,L_f,L_r\right)+\eta_{ij,\,\text{reg.}}^{(5,n)}\left(z,L_f,L_r\right).
\eeq
The coefficients of the leading two powers of logarithms $\log^{5}(1-z)$ and $\log^{4}(1-z)$ of the regular part can be derived using the method of physical evolution kernels of refs.~\cite{Moch:2009hr,Soar:2009yh,deFlorian:2014vta} and agree with our results. 

Furthermore, we investigated the structure of the partonic cross section in the high energy limit. 
The leading logarithmic behaviour of the partonic coefficient function could be computed along the lines of ref.~\cite{Marzani_2009} for the Drell-Yan cross section. To the best of our knowledge, for the $b\bar bH$ cross section this computation currently does not exist. 
However, the structure we observe agrees with our expectation as we observe only a single logarithm at N$^3$LO and the coefficient of this logarithm appears to be universal. Explicitly, we find at NNLO
\bea
&&
\left\{
\eta_{b\bar b}^{(5,2)},\hspace{0.3cm}
\eta_{bg}^{(5,2)},\hspace{0.3cm}
\eta_{bq}^{(5,2)},\hspace{0.3cm}
\eta_{b\bar q}^{(5,2)},\hspace{0.3cm}
\eta_{bb}^{(5,2)},\hspace{0.3cm}
\eta_{gg}^{(5,2)},\hspace{0.3cm}
\eta_{q\bar q}^{(5,2)},\hspace{0.3cm}
\eta_{qg}^{(5,2)}
\right\}\Bigg|_{z^{-1}}\nonumber\\
&&=\left(\frac{\pi ^2}{18}-\frac{13}{54} \right)\times\left\{\frac{2 C_F}{z},\frac{C_A}{z},\frac{C_F}{z},\frac{C_F}{z},\frac{2 C_F}{z},0,0,0\right\},
\eea
and at N$^3$LO
\bea
&&
\left\{
\eta_{b\bar b}^{(5,3)},\hspace{0.3cm}
\eta_{bg}^{(5,3)},\hspace{0.3cm}
\eta_{bq}^{(5,3)},\hspace{0.3cm}
\eta_{b\bar q}^{(5,3)},\hspace{0.3cm}
\eta_{bb}^{(5,3)},\hspace{0.3cm}
\eta_{gg}^{(5,3)},\hspace{0.3cm}
\eta_{q\bar q}^{(5,3)},\hspace{0.3cm}
\eta_{qg}^{(5,3)}
\right\}\Bigg|_{z^{-1}\log(z)}\nonumber\\
&&=C_A \left(-\frac{\zeta_3 }{9}-\frac{\pi ^2 }{54}+\frac{79 }{486} \right)\times\left\{\frac{2 C_F}{z},\frac{C_A}{z},\frac{C_F}{z},\frac{C_F}{z},\frac{2 C_F}{z},0,0,0\right\}.
\eea
\subsection{Analytic results for the partonic coefficient functions}
Our partonic coefficient functions can be expressed in terms of the same set of functions used to represent the results of ref.~\cite{Mistlberger:2018etf}. 
For convenience, we repeat here the most essential definitions.
We define an iterated integral as
\beq
\label{eq:iidef}
J(\vec{\omega},z)=J(\omega_n(z),\dots,\omega_1(z), z)=\int^z_0 dz^\prime \omega_n(z^\prime) J(\omega_{n-1}(z^\prime),\dots,\omega_1(z^\prime),z^\prime)\,.
\eeq
Our partonic coefficient functions can be expressed in terms of linear combinations of the above iterated  integrals with algebraic functions in $z$ as prefactors.
The required integration kernels $\omega_i(z)$ are drawn from the set
\bea
\label{eq:letterset}
\omega_i(z)&\in&\Big\{\frac{1}{1-z},\frac{1}{z},\frac{1}{z+1},1,\frac{1}{\sqrt{z}},\frac{1}{\sqrt{4-z} \sqrt{z}},\frac{\sqrt{z}}{1-z},\frac{1}{\sqrt{z} \sqrt{z+4}},\frac{\sqrt{z}}{\sqrt{z+4}},\frac{1}{\sqrt{4 z+1}},\frac{\sqrt{4 z+1}}{z},\nonumber\\
&&t_{11},t_{12},t_{21},t_{22},\frac{t_{11}}{1-z},\frac{t_{11}}{z},\frac{t_{11}}{z+1},\frac{t_{12}}{1-z},\frac{t_{12}}{z},\frac{t_{12}}{z+1},\frac{t_{21}}{z},\frac{t_{22}}{z}\Big\}.
\eea
The functions $t_{ij}$ are the solutions to the differential equation
\beq
\frac{\partial}{\partial z} \left(\begin{array}{cc} t_{11}(z) & t_{12}(z) \\ t_{21}(z) & t_{22}(z) \end{array}\right) =
\left(
\begin{array}{cc}
 0 &\frac{1}{z}   \\
\frac{3-z}{z^2-11 z-1}  &  \frac{11-2 z}{z^2-11 z-1}\\
\end{array}
\right).
\left(\begin{array}{cc} t_{11}(z) & t_{12}(z) \\ t_{21}(z) & t_{22}(z) \end{array}\right),
\eeq
with
\beq
\left(\begin{array}{cc} t_{11}(1) & t_{12}(1) \\ t_{21}(1) & t_{22}(1) \end{array}\right)= \left(\begin{array}{cc} 0 & 1 \\ 1 & 0 \\\end{array} \right).
\eeq
These functions can be represented in terms of elliptic integrals. 
If an iterated integral only contains integration kernels corresponding to the first three elements of eq.~\eqref{eq:letterset} then it belongs to the class of well known harmonic polylogarithms~\cite{Remiddi:1999ew} (HPLs). More generally, if no integration kernel involving the functions $t_{ij}(z)$ appears, then the iterated integral can be expressed in terms of multiple polylogarithms (MPLs)~\cite{Goncharov1998} evaluated at algebraic arguments.
If also integration kernels involving some $t_{ij}(z)$ appear, the iterated integral cannot be expressed in terms of MPLs alone, but it belongs to a more general class of functions related to elliptic curves. Currently it is unknown if these iterated integrals can be expressed in terms of elliptic multiple polylogarithms~\cite{BrownLevin} or iterated integrals of modular forms~\cite{ManinModular,Brown:mmv}, which have recently appeared in the context of multiloop calculations.
%In case further elements of the set of eq.~\eqref{eq:letterset} appear in an iterated integral then it belongs to a larger class of functions of Goncharov polylogarithms~\cite{Goncharov1998} or functions related to elliptic polylogarithms and iterated integrals of modular forms.
For the purposes we choose to represent our partonic coefficient functions in terms of HPLs and iterated integrals as in eq.~\eqref{eq:iidef}.

\begin{figure*}[!h]
 \begin{center}
\includegraphics[scale=.4]{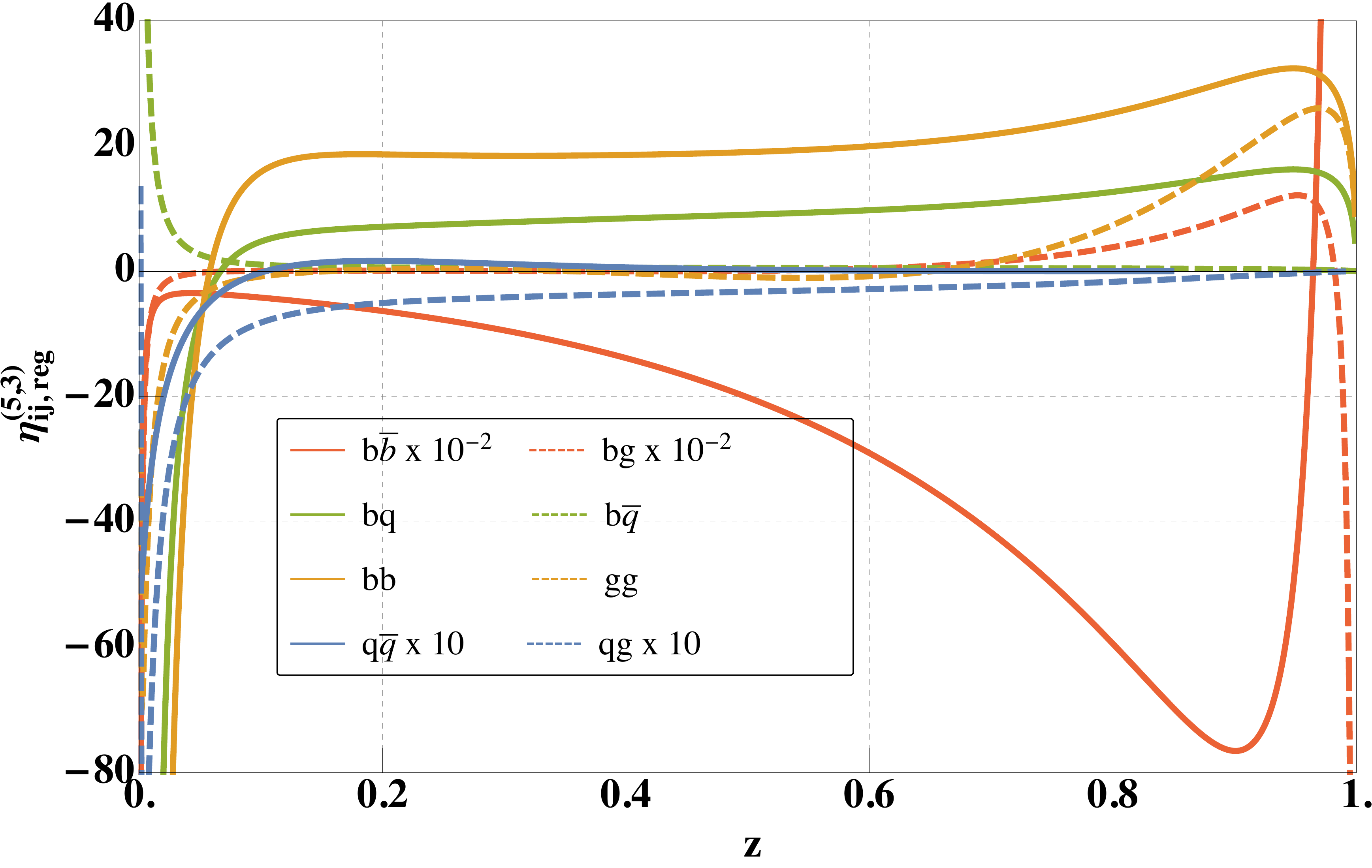}
\end{center}
\caption{\label{fig:PCFRegular} Regular part of the partonic coefficient function at N$^3$LO for all contributing initial state combinations.
Notice, that the $b\bar b$, $bg$ and $qg$ initial state were rescaled uniformly to be visible in the plot.}
\end{figure*}

In order to evaluate the partonic coefficient functions numerically, we find it useful to express them in terms of generalised power series expansions. In ref.~\cite{Mistlberger:2018etf} it was discussed how such iterated integrals relate to one another and how they can be expanded around different numerical points. 
The physical domain for our partonic coefficient functions is given by $z\in[0,1]$. 
By studying the singularities of the functions expressing the partonic coefficients, we can deduce that a generalised power series expansion of the coefficient functions around the point $z=1$ is convergent within the entire physical domain $z\in[0,1]$.
However, in order to reduce the number of terms required to evaluate the partonic coefficient functions to a given numerical accuracy, we choose to expand them around two additional points.
\begin{enumerate}
\item $z\in[\frac{3}{4},1]$: In this interval we expand around the point $z=1$ and define the variable $\bar z=1-z$ for convenience. 
The power series in $\bar z$ is convergent within the entire unit interval but further sub-divisions are desirable in order to avoid loss of numerical accuracy when including only few orders in the expansion.
We provide 50 terms in the series expansion around $\bar z=0$.
\item $z\in[\frac{1}{13},\frac{3}{4}]$: Within this interval we expand around the point $z=\frac{1}{2}$ and define the variable $w=\frac{1}{2}-z$ for convenience. 
We provide 200 terms in the expansion around $w=0$. 
Formally, this expansion around $w=0$ is convergent in the entire interval $z\in[0,1]$.
\item $z\in[0,\frac{1}{13}]$: In this interval we expand our partonic coefficient functions around the point $z=0$ and we provide 100 terms in this expansion. 
Contrary to the previous two expansions, this one is only convergent within the interval $z\in[0,-\frac{1}{2} \left(11-5 \sqrt{5}\right)]$.
\end{enumerate}
With the provided number of terms in the different series expansions the partonic coefficient functions can be evaluated with a relative numerical precision of at least $10^{-10}$.
While the formal radius of convergence of the different expansions listed above refers to the validity of the expansions, we advise to stick to the suggested intervals in order to achieve a numerical accuracy of the partonic coefficient function of at least ten significant digits.
We provide digital files containing the partonic coefficient functions through N$^3$LO as ancillary material of this article. Figure~\ref{fig:PCFRegular} shows the individual regular partonic coefficient functions for the eight different partonic initial states. 

% !TEX root = paper.tex
\section{Phenomenological results in the five-flavour scheme}
\label{sec:5flav}
In this section we analyse the impact of the N$^3$LO corrections Higgs boson production cross section via bottom quark fusion in the 5FS.
We work with a Higgs mass of $m_H=125.09\textrm{ GeV}$ and the pole mass of the bottom quark is $m_b=4.58\textrm{ GeV}$. 
The strong coupling and the Yukawa coupling $y_b$ are evaluated at the renormalisation scale $\mu_R^2$ using three-loop running in the $\overline{\textrm{MS}}$-scheme~\cite{Tarasov:1980au,Larin:1993tp,vanRitbergen:1997va,Baikov:2016tgj,Herzog:2017ohr,Harlander:2003ai,Chetyrkin:1997dh,Czakon:2004bu,Baikov:2014qja}, and we start the evolution from $\alpha_s(m_Z^2)=0.118$ and $m_b(m_b)=4.18\textrm{ GeV}$. 

We use the \texttt{PDF4LHC15\_nnlo\_mc} set~\cite{Butterworth:2015oua} parton distribution functions if not stated otherwise explicitly.
Throughout this article we only consider contributions proportional to $\mathcal{O}(y_b^2)$. We however remind the reader that bottom quark fusion contributions proportional to $\mathcal{O}(y_by_t)$ and $\mathcal{O}(y_t^2)$ are relevant as already discussed in refs.~\cite{Wiesemann:2014ioa,Deutschmann:2018avk}.

\subsection{Dependence on the perturbative scales}

\begin{figure*}[!ht]
\centering
\includegraphics[width=0.49\textwidth]{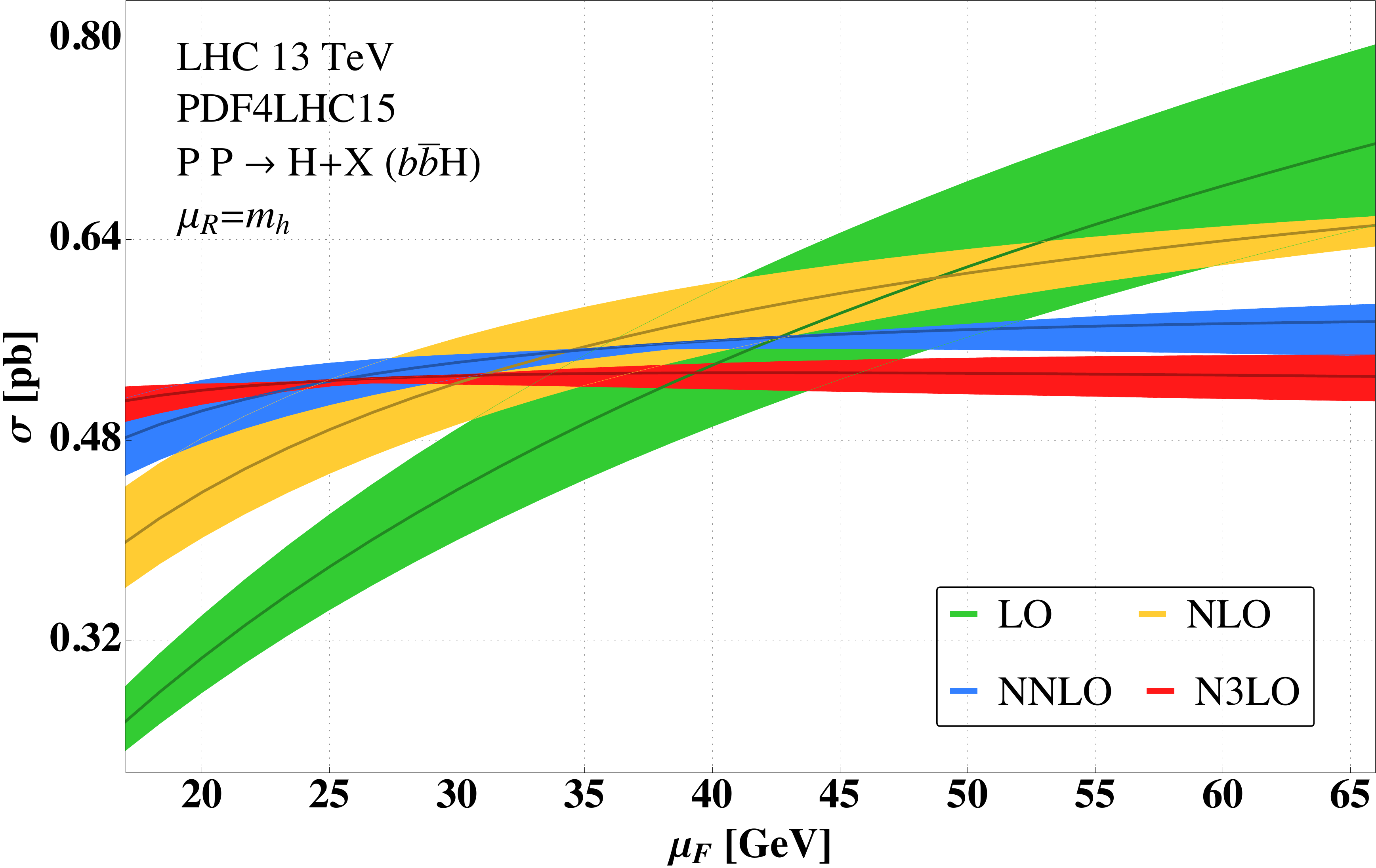}
\includegraphics[width=0.49\textwidth]{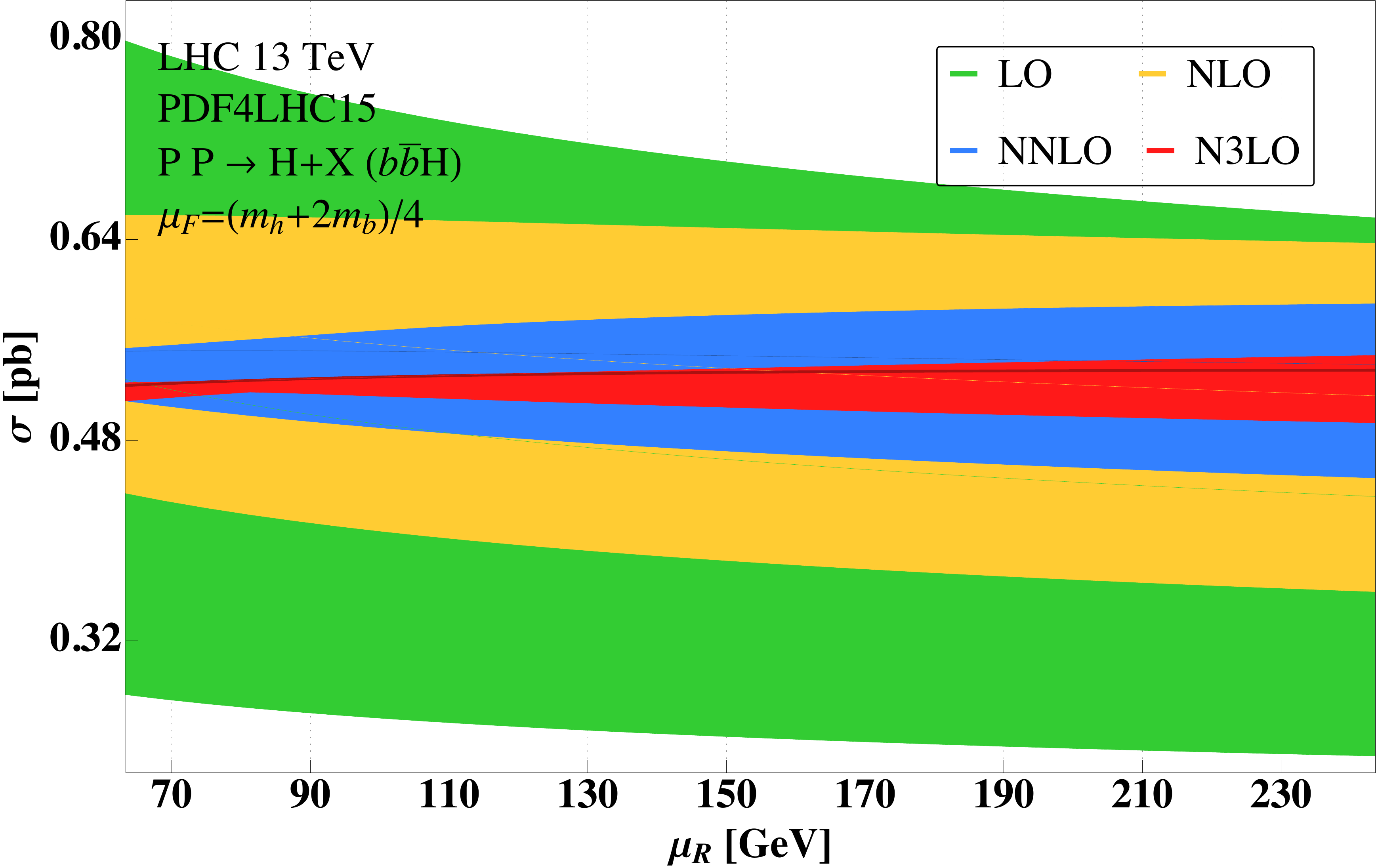}
\caption{\label{fig:scalevariation}
Variation of the $b\bar b\rightarrow H+X$ cross section with the factorisation scale $\mu_F$ (left) and renormalisation scale $\mu_R$ (right). 
The bands in the left (resp. right) panel indicate the range of the variation of the prediction when modifying the factorisation scale $\mu_F$ (resp. $\mu_R$) by the two factors $\frac{1}{2}$ and $2$.
 Predictions in green, yellow, blue and red correspond to LO, NLO, NNLO and N$^3$LO respectively.
}
\end{figure*}

Through N$^3$LO our cross section is independent of the factorisation and renormalisation scales.
However, the numerical values for cross section predictions will vary depending on the choice of the values for the perturbative scales since the evolution of the PDFs, the strong coupling and the Yukawa coupling are performed in a resummed fashion.
At NLO it was argued in refs.~\cite{Boos:2003yi,Plehn:2002vy,Rainwater:2002hm,Maltoni:2003pn} that the $t$-channel singularity in the gluon-initiated process $gb\to bH$ leads to a collinear logarithm of the form $\log(4\mu_F/m_H)$ in the inclusive cross section and that consequently a low value for the factorisation scale should be preferred. 
In refs.~\cite{Harlander:2003ai,Harlander:2012pb,deFlorian:2016spz} it was observed that choosing low factorisation scales leads to faster stabilisation of the perturbative series.
We consequently follow this approach and choose the as the central values for our perturbative scales:
\beq
\label{eq:scalechoice}
\mu_R^{\text{cent.}}=m_H,\hspace{1cm} \mu_F^{\text{cent.}}=\frac{m_H+2m_b}{4}\,.
\eeq
Figure~\ref{fig:scalevariation} shows the dependence of the hadronic cross section on the factorisation (left) and renormalisation (right) scales. 
The bands in the two figures are obtained by varying \emph{one} particular scale up and down by a factor of two around the central value.
We observe in fig.~\ref{fig:scalevariation} that including higher-order perturbative corrections reduces the dependence of the hadronic cross section on both perturbative scales since the span of the bands is reduced by the inclusion of higher-order corrections.
We also notice that the perturbative series is relatively well behaved for low values of the factorisation scale. This strengthens the case for our choice of central value for the factorisation scale.

\begin{figure*}[!t]
\centering
\begin{subfigure}[t]{0.49\textwidth}
\hspace{0.01cm}
\includegraphics[width=0.96\textwidth]{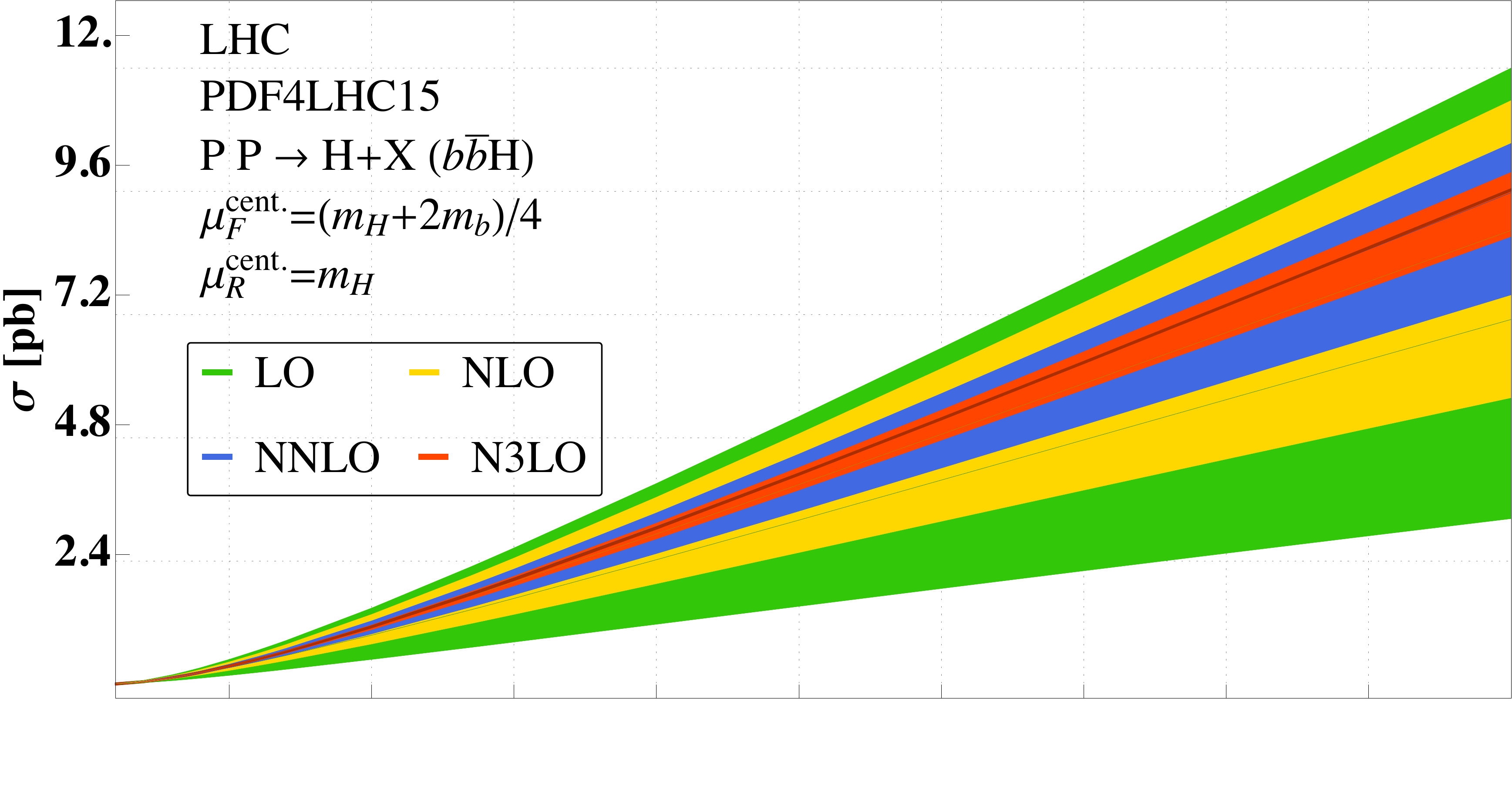}\vspace{-0.4cm}\\
\includegraphics[width=\textwidth]{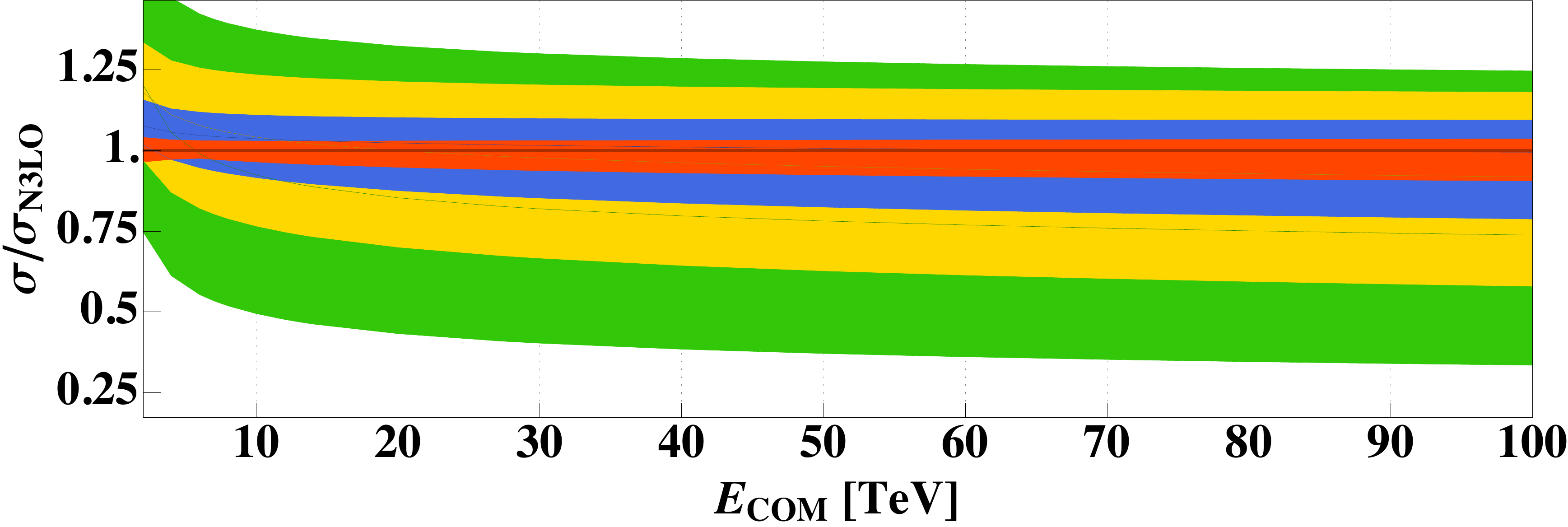}
\end{subfigure}
\begin{subfigure}[t]{0.49\textwidth}
\hspace{0cm}
\includegraphics[width=0.96\textwidth]{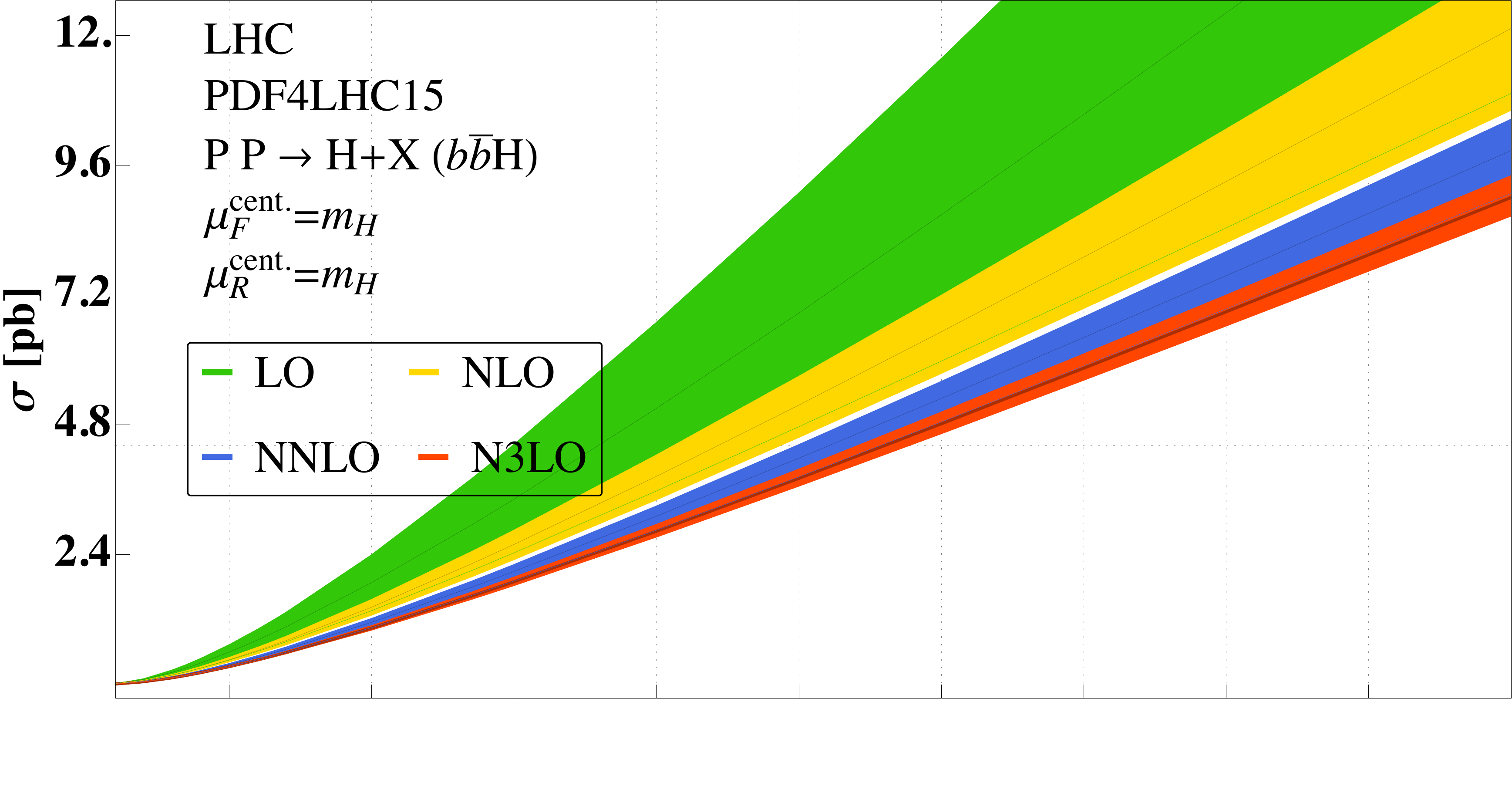}\vspace{-0.4cm}\\
\includegraphics[width=\textwidth]{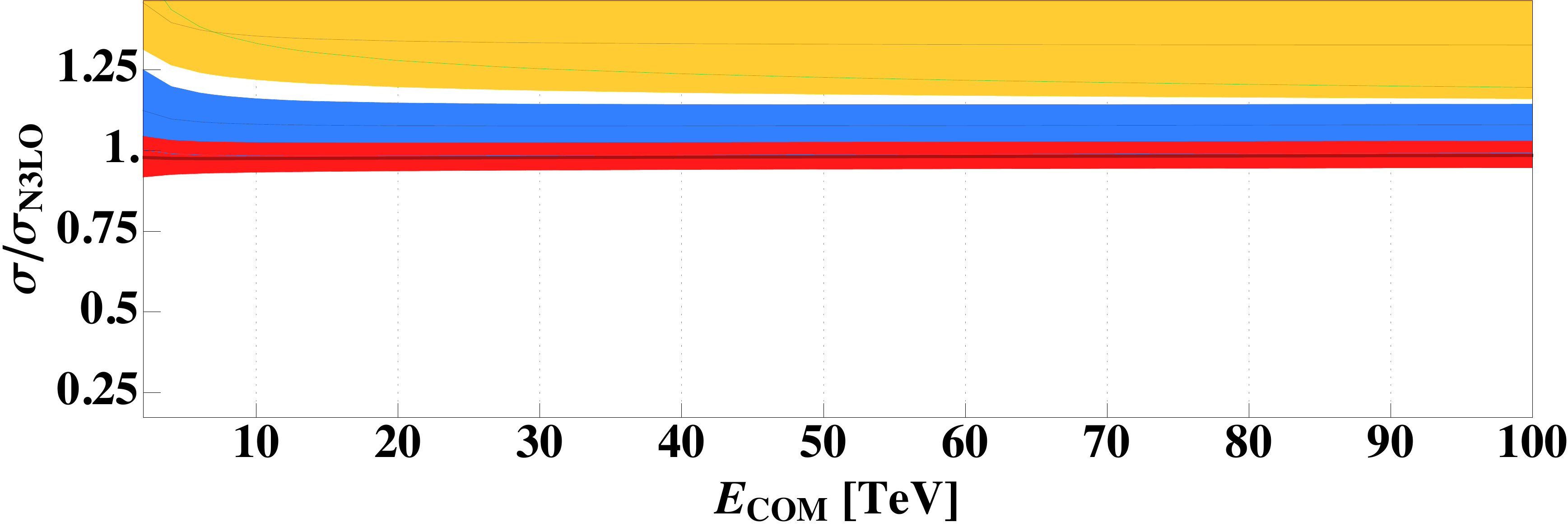}
\end{subfigure}
\caption{\label{fig:energyvariation}
The hadronic cross section as a function of the collider energy. Green, orange, blue and red bands correspond to predictions through LO, NLO, NNLO and N$^3$LO respectively. 
The left figure shows predictions with $\mu_F^{\text{cent.}}=(m_H+2 m_b)/4$ and the right figure with $\mu_F^{\text{cent.}}=m_H$.
The bottom panel of both pictures shows the cross section predictions normalised to the N$^3$LO prediction with $\mu_F^{\text{cent.}}=(m_H+2 m_b)/4$.
}
\end{figure*}

Figure~\ref{fig:energyvariation} shows the cross section for the production of a Higgs boson in bottom quark fusion for various hadron collider energies. 
Different colours refer to different orders of the perturbative expansion, and the
bands correspond to varying the perturbative scales by a factor of two around their central value while satisfying the inequality (7-point variation)
\beq
\label{eq:scaleconstraint}
\frac{1}{2}\leq \frac{\mu_R/ \mu_R^{\text{cent.}} }{\mu_F/ \mu_F^{\text{cent.}} }\leq 2\,.
\eeq
We observe a reduction of the size of the scale variation bands when including higher-order corrections.
%, which we interpret as a more precise prediction for the inclusive $bbH$ cross section.
The left panel of fig.~\ref{fig:energyvariation} displays the cross section with the central scale choice of eq.~\eqref{eq:scalechoice}. 
The right panel of fig.~\ref{fig:energyvariation} shows the same quantity but with the choice $\mu_F^{\text{cent.}}=m_H$. 
We find that the nominal value of the cross section at N$^3$LO is comparable for these two choices. 
However, the perturbative corrections are much larger in the latter case, thus further supporting our choice of a low factorisation scale for this process.

\subsection{PDF and $\alpha_s$ uncertainties}
We take the PDFs and the strong coupling constant as external input. 
These quantities are naturally associated with an uncertainty that we asses following the guidelines of the providers of these quantities.
In particular, we use the  \texttt{PDF4LHC15\_nnlo\_mc}  set~\cite{Butterworth:2015oua} as our default PDF set and follow the Monto-Carlo prescription outlined in ref.~\cite{Butterworth:2015oua} in order to determine the PDF uncertainty of our cross section.
In particular, following this prescription the hadronic cross section is computed with 100 different PDF sets and the resulting values are then ordered by nominal size. 
The PDF uncertainty is then determined by 
\beq
\delta(\textrm{PDF})=\pm\frac{\sigma^{(5)}_{84}-\sigma^{(5)}_{16}}{\sigma^{(5)}_{84}+\sigma^{(5)}_{16}},
\eeq
where, the $\sigma^{(5)}_{i}$ corresponds to the $i^{\text{th}}$ member of the ordered set. 
As a central value for cross section predictions is recommended to be 
\beq
\label{eq:PDFMean}
\bar\sigma^{(5)} =\frac{1}{2}\left(\sigma^{(5)}_{84}+\sigma^{(5)}_{16}\right).
\eeq

\begin{figure*}[!t]
\centering
\includegraphics[width=0.49\textwidth]{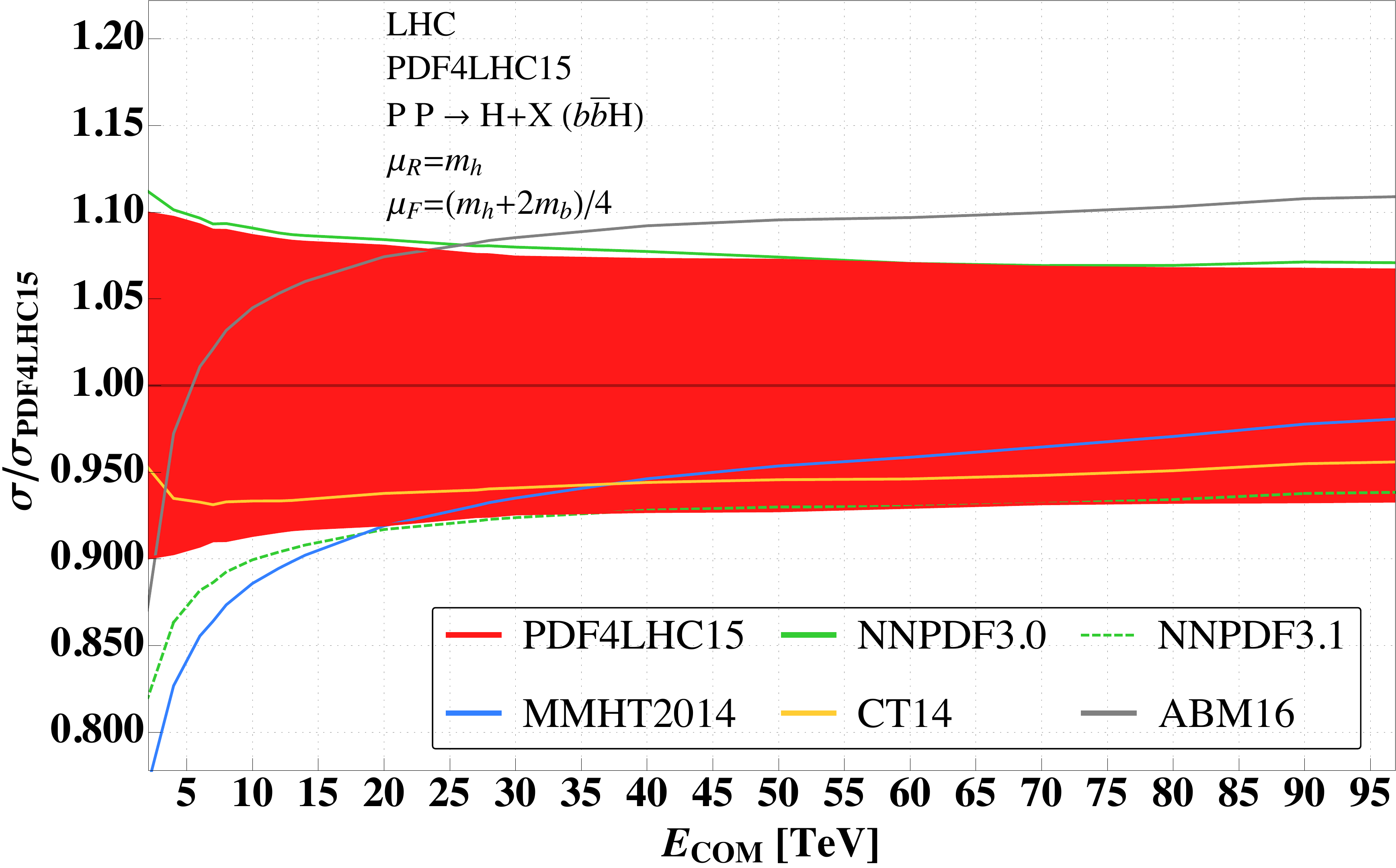}
\includegraphics[width=0.49\textwidth]{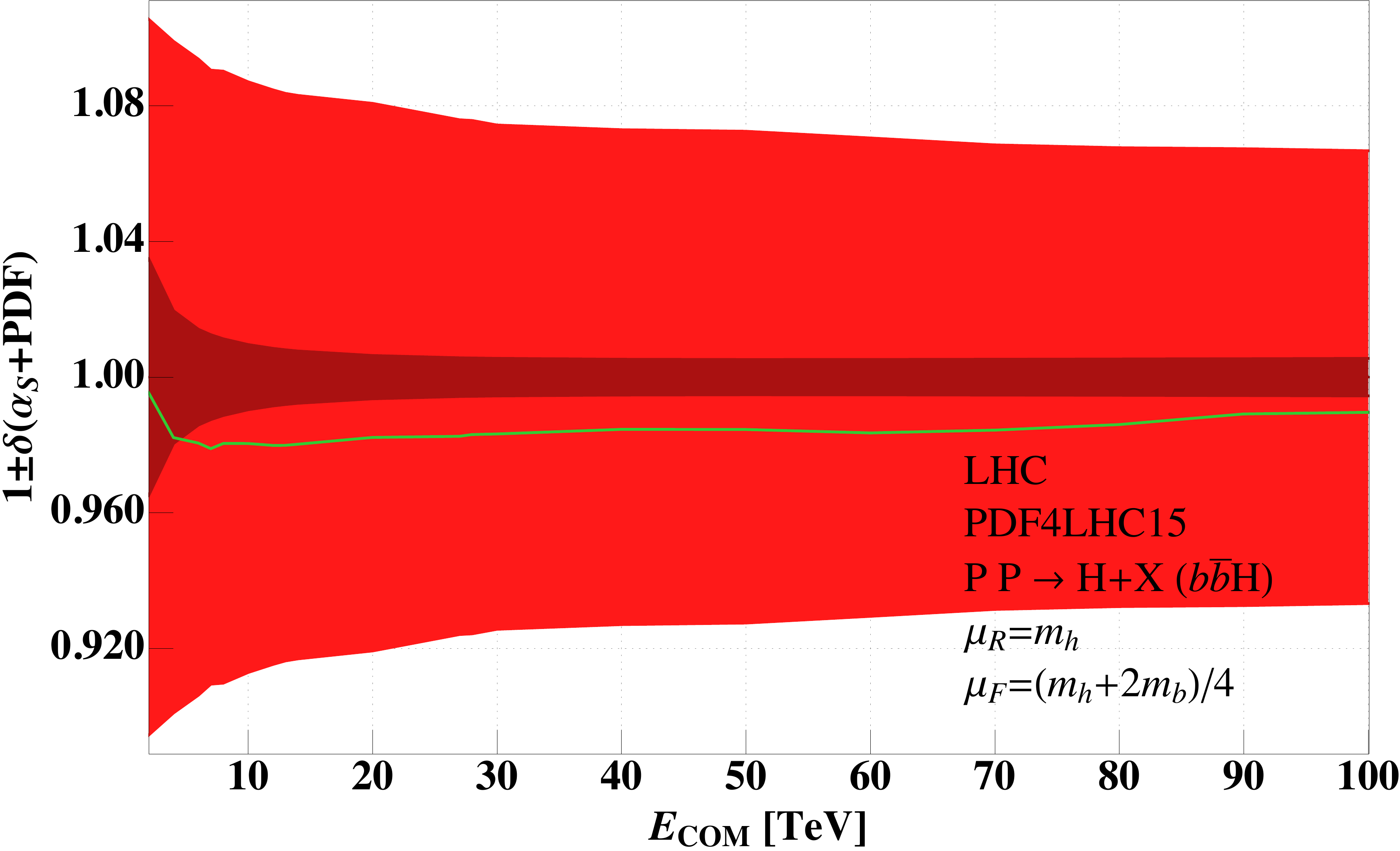}
\caption{\label{fig:PDFError}
Dependence of the cross section on the choice of PDF as a function of the energy normalised to the central value computed according to eq.~\eqref{eq:PDFMean}. 
On the left the red band shows the uncertainty computed with the PDF4LHC15 Monte-Carlo prescription and the lines correspond to predictions obtained with other PDF sets. 
On the right the dark and light red bands correspond to $\delta(\alpha_S)$ and $\delta(\alpha_S+\textrm{PDF})$ respectively.
The green line on the right is the ratio of the prediction obtained with the central PDF set of PDF4LHC15 to the central value obtained according to eq.~\eqref{eq:PDFMean}.
}
\end{figure*}
Figure~\ref{fig:PDFError} shows the resulting PDF uncertainty as a function of the collider energy. 
Furthermore, we compare different PDF sets with prediction based on the PDF4LHC15 set. 
In particular we study the sets
\begin{itemize}
\item \texttt{CT14nnlo\_as\_0118}~\cite{Dulat:2015mca}\,,
\item \texttt{MMHT2014nlo68clas118}~\cite{Harland-Lang:2014zoa}\,,
\item \texttt{ABMP16\_5\_nnlo}~\cite{Alekhin:2016uxn}\,,
\item \texttt{NNPDF30\_nnlo\_as\_0118}~\cite{Ball:2014uwa}\,,
\item \texttt{NNPDF31\_nnlo\_as\_0118}~\cite{Ball:2017nwa}\,.
\end{itemize}
We observe a sizable PDF uncertainty from $7-9\%$. 
Comparing the predictions based on the PDF4LHC15 set with the other PDF set we see significant differences. 
The PDF4LHC15 set itself is a statistical combination of the CT14, MMHT and NNPDF3.0 sets, and we observe in fig.~\ref{fig:PDFError} that indeed the resulting prediction is in between the three input sets.
NNPDF3.1 is an updated version of NNPDF3.0 and technically supersedes the latter. 
Consequently, it is possible that a combination of CT14, MMHT and NNPDF3.1 into an updated version of a PDF4LHC combination would lead to a significantly lower central prediction of the $b\bar b H$ cross section.
However, such a  study is beyond the scope of this article.

In order asses the uncertainty due to the imprecise knowledge of the strong coupling constant, the authors of  ref.~\cite{Butterworth:2015oua}  provide two PDF sets within the    \texttt{PDF4LHC15\_nnlo\_mc\_pdfas}   set that allow to vary the strong coupling constant by $\pm 0.0015$ in a correlated fashion.
The associated uncertainty is computed as 
\beq
\delta(\alpha_S)=\pm\frac{1}{2\bar \sigma^{(5)}}\left|\sigma^{(5)}(\alpha_S=0.1195)-\sigma^{(5)}(\alpha_S=0.1165)\right|.
\eeq
Following the recommendation of ref.~\cite{Butterworth:2015oua}  this uncertainty can then be combined in quadrature with the PDF uncertainty:
\beq
\delta(\alpha_S+\textrm{PDF})=\sqrt{\delta(\textrm{PDF})^2+\delta(\alpha_S)^2}.
\eeq

The definition of the value for the prediction of the inclusive cross section in eq.~\eqref{eq:PDFMean} can be compared with the prediction that is obtained with the central member of the  \texttt{PDF4LHC15\_nnlo\_mc}  set. 
Their ratio is shown in fig.~\ref{fig:PDFError} in green on the right. 
While there is a non-negligible difference the two predictions are compatible within the PDF uncertainties.

\subsection{PDF theory uncertainty}
PDFs are currently determined using NNLO cross sections as input for their extraction from a wide set of measurements.
Consequently, we refer to these PDFs as NNLO PDFs. 
Since our cross section is computed at N$^3$LO this leads to a mismatch that can ultimately be remedied by using N$^3$LO cross sections for the PDF extraction. 
In the meantime we estimate the potential impact of this mismatch on our cross section predictions. 
In ref.~\cite{Anastasiou:2016cez} a prescription was introduced that studies the variation of the NNLO cross section as NNLO or NLO PDFs are used. 
This defines the PDF theory uncertainty
\beq
\label{eq:PDFTH}
\delta(\text{PDF-TH})=\frac{1}{2}\left|\frac{\sigma^{\text{NNLO, NNLO-PDFs}}-\sigma^{\text{NNLO, NLO-PDFs}}}{\sigma^{\text{NNLO, NNLO-PDFs}}}\right|.
\eeq
Here, the factor $\frac{1}{2}$ is introduced as it is expected that this effect becomes smaller at N$^3$LO compared to NNLO.

\begin{figure*}[!t]
\centering
\begin{subfigure}[t]{0.49\textwidth}
\includegraphics[width=\textwidth]{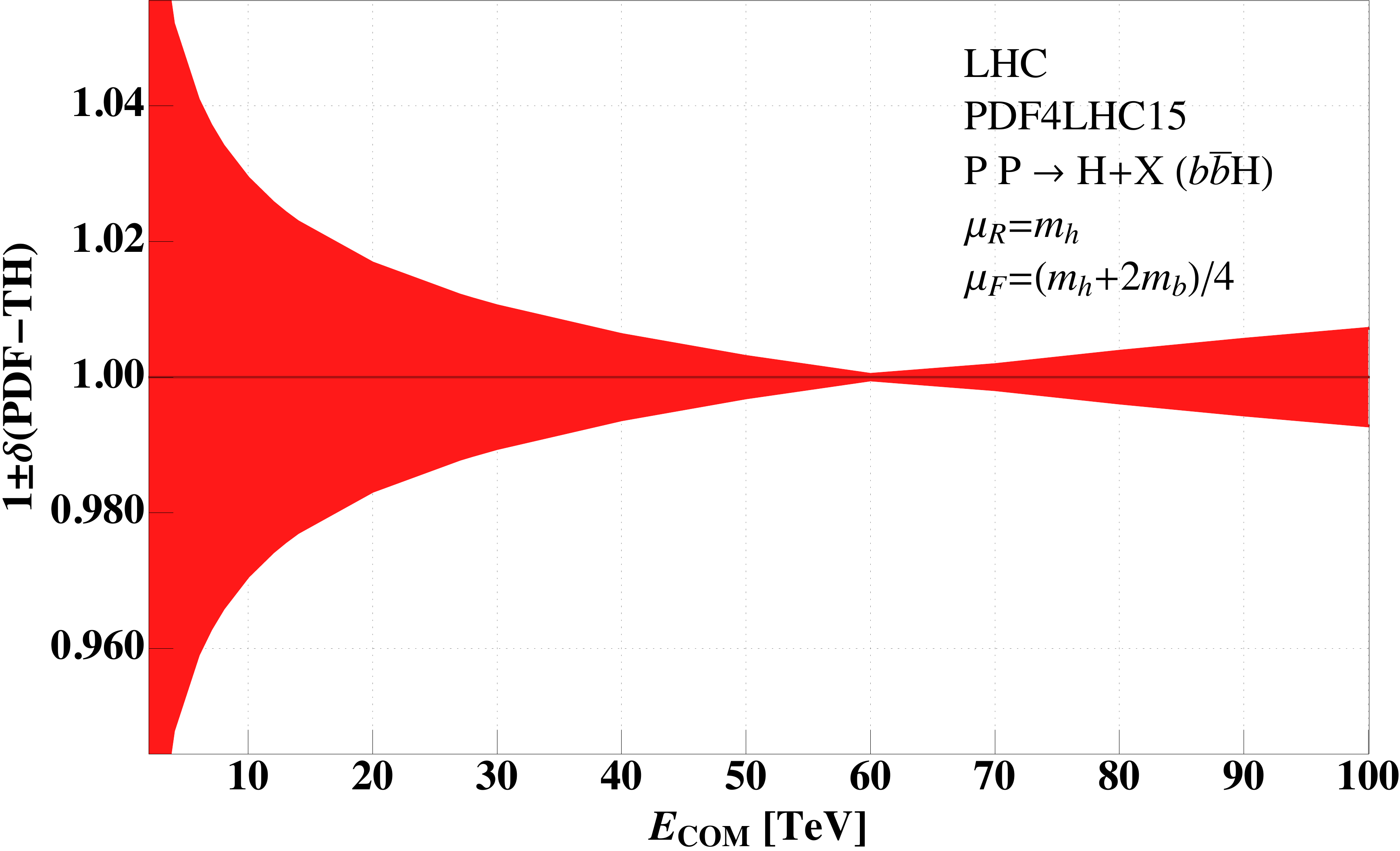}
\subcaption{
\label{fig:PDFTHError}
$\delta(\text{PDF-TH})$ uncertainty as a function of the collider energy.
}
\end{subfigure}
\begin{subfigure}[t]{0.49\textwidth}
\includegraphics[width=\textwidth]{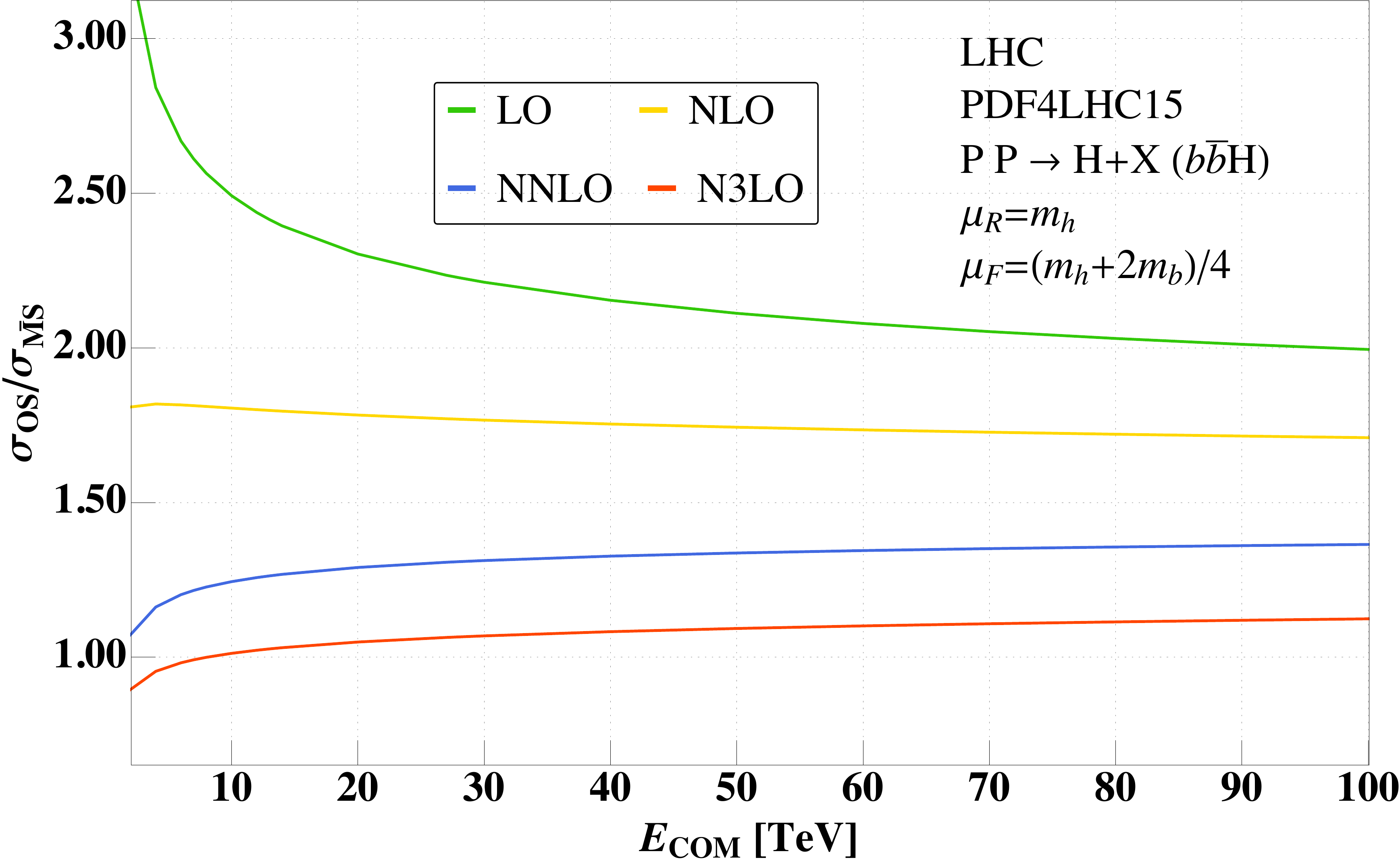}
\subcaption{
\label{fig:OSMass}
Ratio of $\sigma^{(5)}$ computed with an on-shell bottom quark mass over the same with an $\overline{\text{MS}}$ mass as a function of the collider energy.
}
\end{subfigure}
\caption{Impact of missing N$^3$LO PDFs and bottom quark mass schemes on the $b\bar b H$ cross section.}
\end{figure*}
Figure~\ref{fig:PDFTHError} displays $\delta(\text{PDF-TH})$ as a function of the collider energy. 
Throughout this uncertainty is smaller than the PDF uncertainty. 
We interpret the numerical crossing point at about $60$ TeV as a coincidence and a simple consequence of the method we use to estimate this uncertainty. 
Consequently, this does not mean that there is no PDF theory uncertainty for a 60 TeV collider and we assign always at least a $1\%$ uncertainty whenever the prescription of eq.~\eqref{eq:PDFTH} falls below.

\subsection{Bottom quark mass uncertainty}

According to the PDG~\cite{Tanabashi:2018oca} the bottom quark mass in the $\overline{\text{MS}}$-scheme is determined to be 
\beq
m_b(m_b)=4.18^{+0.03}_{-0.02} \text{ GeV}.
\eeq
Since the cross section in the 5FS is proportional to the square of the bottom quark mass the hadronic $b\bar bH$ cross section is affected by the corresponding uncertainty
\beq
\delta(m_b)=\frac{(\delta m_b)^2}{m_b^2}={}^{+1.44 \%}_{-0.95 \%}.
\eeq

The bottom quark mass evaluated at the renormalisation scale is completely factorised from the partonic coefficient functions as can be seen in eq.~\eqref{eq:sigma_5FS}.  
We perform the scale evolution via a numerical solution to the evolution equation using anomalous dimensions at $(n+1)$ perturbative order in order to compute the N$^n$LO cross section:
\beq
\frac{\partial}{\partial \log\mu^2} m_b(\mu^2) =m_b(\mu^2)\sum\limits_{i=0}^\infty a_s(\mu^2)^i \gamma^{(i)}.
\eeq
The constants $\gamma^{(i)}$ are taken from ref.~\cite{Vermaseren:1997fq}. 
Overall, we find that truncating the anomalous dimension at the $(n+1)^{\text{th}}$ order slightly improves the rate of convergence of the perturbative expansion. 
However, we find that the value of $m_b(\mu_R^{\text{cent.}}=m_H)$ changes at the sub-permille level if we are using three-loop or four-loop anomalous dimensions, cf. tab.~\ref{tab:mb}. Consequently, we do not assign an additional uncertainty for the exact implementation of the bottom quark mass.
\begin{table}[!h]
\begin{center}
\begin{tabular}{c | c | c | c | c | c}
\hline\hline
&  $\gamma^{(n>-1)}=0$ & $\gamma^{(n>0)}=0$ & $\gamma^{(n>1)}=0$ & $\gamma^{(n>2)}=0$ & $\gamma^{(n>3)}=0$\\
\hline
$m_b(m_h)\, [\text{GeV}]$ & 4.18 & 3.01 & 2.81 & 2.79 & 2.79 \\
\hline\hline
\end{tabular}
\caption{\label{tab:mb}The running of the bottom quark mass in the $\overline{\textrm{MS}}$-scheme at different orders.}
\end{center}
\end{table}

Alternatively to the $\overline{\text{MS}}$-scheme, we derive predictions for the $b\bar b H$ cross section using the on-shell bottom quark mass.
Using the three-loop conversion relation of refs.~\cite{Chetyrkin:1999qi,Marquard:2016dcn} we find that the on-shell bottom quark mass is given by
\beq
m_b^{\text{OS}}(m_b(m_b))=4.92 \textrm{ GeV}\,.
\eeq

Figure~\ref{fig:OSMass} shows the ratio of the $b\bar b H$ cross section with computed with on-shell bottom quark mass at different perturbative orders to the same computed with $\overline{\text{MS}}$ mass at N$^3$LO. 
We observe that as the perturbative order is increased the predictions based on different mass schemes approach each other. 
However, the perturbative convergence of the cross section predictions using on the on-shell mass is quite slow.
In part this can be attributed to the fact that we are not resumming the mass evolution as in the $\overline{\text{MS}}$-scheme. 
At LO the bottom quark mass in eq.~\eqref{eq:sigma0} is now evaluated with its on-shell value and the ratio of the normalisation factors $\hat\sigma_0$ of the two different schemes is $\sim 2.67$.
Furthermore, it is well known that the conversion from $\overline{\text{MS}}$ to on-shell scheme is affected by large perturbative corrections (see for example refs.~\cite{Chetyrkin:1999qi,Marquard:2016dcn}).
Based on the above observations we recommend the treatment of the bottom quark mass as in our default set-up.

%\input{PDFError.tex}
% !TEX root = paper.tex

\section{The FONLL matching procedure}
\label{sec:fonnll}

In order to have precise theoretical predictions it is desirable to combine the 4FS and 5FS into a single prediction which retains finite mass effects through a certain order in perturbation theory while at the same time resumming the collinear logarithms to all orders in the strong coupling. 
Various methods have been proposed in the literature to combine the two schemes~\cite{Harlander:2011aa,Bonvini:2015pxa,Bonvini:2016fgf,Forte:2015hba,Forte:2016sja}. Here we focus on the so-called FONLL scheme, first introduced in refs.~\cite{Cacciari:1998it,Forte:2010ta} for hadron production in hadronic collisions and deep inelastic scattering and recently applied to Higgs~\cite{Forte:2015hba,Forte:2016sja} and $Z$-boson~\cite{Forte:2018ovl} production in bottom quark fusion in proton collisions. The original versions of refs.~\cite{Forte:2015hba,Forte:2016sja}, however, contained some misprints, and we therefore reproduce all formulas here for completeness.

At all perturbative orders, the cross sections in the 4FS and 5FS in eqs.~\eqref{eq:sigma_4FS} and~\eqref{eq:sigma_5FS} are identical up to power suppressed terms (and possibly up to non-perturbative effects encoded in the different PDFs),
\beq
\sigma^{(4)} - \sigma^{(5)} = \ord\left({m_b^2}\right)\,.
\eeq
A similar relation, however, does not hold at the level of the partonic coefficient functions calculated in the two schemes. Indeed, the coefficient functions in the 4FS develop logarithmic divergencies in the limit of a vanishing bottom quark mass, which are not captured by the coefficient functions in the 5FS. Instead, these $m_b$-dependent logarithms are encoded (and resummed) into the PDFs and the strong coupling constant in the 5FS.

The starting point of the FONLL method is to express both computations in terms of a common set of PDFs and $\alpha_s$, namely the ones in the 5FS. The relation between the strong coupling constant and the PDFs in the two schemes takes the form,
\beq\bsp\label{eq:5_to_4}
\alpha_s^{(5)}(\mu_R^2) &\,= \alpha_s^{(4)}(\mu_R^2) + \sum_{n=1}^\infty c_n(\mu_R^2/m_b^2)\,\alpha^{(4)}_s(m_b^2)\,,\\
f_i^{(5)}(x,\mu_F^2) &\, = \sum_{j=-4}^4K_{ij}(x,L_b,\alpha_s^{(4)}(\mu_F^2))\otimes f_j^{(4)}(x,\mu_F^2)\,, \qquad -5\le i\le 5\,,
\esp\eeq
with $L_b\equiv \log\frac{\mu_F^2}{m_b^2}$. The explicit form of the kernels $K_{ij}$ relevant here can be obtained from ref.~\cite{Buza:1996wv}. In particular, they have the property that $K_{ij} = \delta_{ij}\,\delta(1-x)+\ord(\alpha_s)$ for $|i|\neq 5$ and $K_{ij} = \ord(\alpha_s)$ for $i=\pm 5$. This allows us to invert eq.~\eqref{eq:5_to_4} order by order in the coupling, and to express the cross section in the 4FS in eq.~\eqref{eq:sigma_4FS} in terms of the coupling and the PDFs in the 5FS,
\beq\label{eq:sigma_4FS_PDF5}
\sigma^{(4)} = \tau\sum_{i,j=-4}^4\,\mathscr{L}_{ij}^{(5)}(\tau,\mu_F^2)\otimes {B}_{ij}\!\left(\tau,L_f,L_r,{m_b^2},\alpha_s^{(5)},y_b^2\right)\,,
\eeq
where the partonic coefficient functions admit the perturbative expansion:
\beq\bsp
{B}_{ij}\!\left(z,L_f,L_r,{m_b^2},\alpha_s^{(5)},y_b^2\right)=\hat\sigma_0(y_b^2,m_H^2) \sum_{n=2}^\infty a_s^{(5)n}\,{B}_{ij}^{(n)}\!\left(z,L_f,L_r,{m_b^2}\right)\,.
\esp\eeq
Through third order in the strong coupling, the relation between the partonic coefficient functions in eqs.~\eqref{eq:sigma_4FS} and~\eqref{eq:sigma_4FS_PDF5} reads,
\beq\bsp
\label{eq:4FSterms}
B_{gg}^{(3)}\!\left(z,L_f,L_r,{m_b^2}\right) &\,= 
\eta_{gg}^{(4,3)}\!\left(z,L_f,L_r,{m_b^2}\right) + \frac{2}{3}\,T_f\,L_r\,\eta_{gg}^{(4,2)}\!\left(z,L_f,L_r,{m_b^2}\right)\,,\\
B_{q\bar{q}}^{(3)}\!\left(z,L_f,L_r,{m_b^2}\right) &\,= 
\eta_{q\bar{q}}^{(4,3)}\!\left(z,L_f,L_r,{m_b^2}\right) + \frac{2}{3}\,T_f\,(L_r-L_b)\,\eta_{q\bar{q}}^{(4,2)}\!\left(z,L_f,L_r,{m_b^2}\right)\,,
\esp\eeq
with $T_f=\frac{1}{2}$.

By inserting the expression of the PDFs in the 4FS in terms of those in the 5FS back into eq.~\eqref{eq:5_to_4}, we can re-express the $b$-PDF entirely in terms of the PDFs for the other parton flavours in the 5FS. Through the order we need it, this relation reads
\begin{align}\label{eq:bto5_exp}
&f_b^{(5)}(x,\mu_F^2) = f_{\bar{b}}^{(5)}(x,\mu_F^2)= a_s^{(5)}(\mu_F^2)\,\cA_{bg}^{(1)}(x,L_b)\otimes f_g^{(5)}(x,\mu_F^2)\\
\nonumber&\,+a_s^{(5)}(\mu_F^2)^2\,\Bigg[\cA_{bg}^{(2)}(x,L_b)\otimes f_g^{(5)}(x,\mu_F^2) + \cA_{b\Sigma}^{(2)}(x,L_b)\otimes \sum_{\substack{i=-4\\ i\neq 0}}^4f_i^{(5)}(x,\mu_F^2)\Bigg]+\ord(a_s^{(5)}(\mu_F^2)^3)\,.
\end{align}
Note that the bottom and anti-bottom distributions are only identical through the first two orders in the strong coupling constant, and they will start to differ starting from $\ord(a_s^{3})$ (cf., e.g., ref.~\cite{Catani:2004nc}). The kernels $\cA_{bg}^{(k)}$ and $\cA_{b\Sigma}^{(2)}$ can be found in ref.~\cite{Buza:1996wv}.
Inserting this relation into eq.~\eqref{eq:sigma_5FS}, we can write the cross section in the 5FS as $\sigma^{(4-5)}$ in a way that does not involve the $b$-PDF and which is formally equal to $\sigma^{(5)}$ up to third order in $\alpha_s^{(5)}$,
\beq\label{eq:sigma_4-5}
\sigma^{(4-5)} = \tau\sum_{i,j=-4}^4\,\mathscr{L}_{ij}^{(5)}(\tau,\mu_F^2)\otimes {A}_{ij}\!\left(\tau,L_f,L_r,L_b,\alpha_s^{(5)},y_b^2\right)\,.
\eeq
The partonic coefficient functions ${A}_{ij}$ can be expressed in terms of the partonic coefficient functions in the 5FS in eq.~\eqref{eq:sigma_5FS} and the kernels in eq.~\eqref{eq:bto5_exp}. In the following we only show this relation for $\mu_R=\mu_F$, and we suppress the dependence of all functions on their arguments for readability. 
If we denote the coefficient of $y_b^2(\mu_F^2)\,a_s(\mu_F)^n$ by ${A}_{ij}^{(n)} = \eta_{ij}^{(5,n)} + \delta\eta_{ij}^{(5,n)}$, we have
\beq\bsp
\delta\eta_{gg}^{(5,2)}&\, =  4\,\cA^{(1)}_{bg}\otimes{\eta}_{gb}^{(5,1)} + 2\,\cA^{(1)}_{bg}\otimes\cA^{(1)}_{bg}\,,\\
\delta\eta_{gg}^{(5,3)}&\, =   4\,\cA^{(2)}_{bg}\otimes{\eta}_{gb}^{(5,1)} + 4\,\cA^{(1)}_{bg}\otimes\hat{\eta}_{gb}^{(2)}+4\,\cA^{(1)}_{bg}\otimes\cA^{(2)}_{bg} +2\,\cA^{(1)}_{bg}\otimes\cA^{(1)}_{bg}\otimes{\eta}_{b\bar{b}}^{(5,1)}\,,\\
\delta\eta_{gq}^{(5,3)}&\, =\delta\eta_{g\bar{q}}^{(5,3)} =  2\,\cA_{b\Sigma}^{(2)}\otimes{\eta}_{gb}^{(5,1)}+\cA^{(1)}_{bg}\otimes{\eta}_{bq}^{(5,2)}+\cA^{(1)}_{bg}\otimes{\eta}_{b\bar{q}}^{(5,2)}+2\,\cA^{(1)}_{bg}\otimes\cA_{b\Sigma}^{(2)}\,,
\esp\eeq
while $\delta\eta_{ij}^{(5,2)} = \delta\eta_{ij}^{(5,3)} = 0$ for all other channels. We have performed all these convolutions analytically using the {\sc PolyLogTools} package~\cite{Duhr:2019tlz}.  The analytic expressions for the convolutions in terms of multiple polylogarithms are provided as ancillary material with the arXiv submission.
%The results for the non-vanishing coefficient functions $\delta\eta_{ij}^{(5,n)}$ for $n=2,3$ are presented in Appendix~\ref{app:matching_coefficients}.

Using these definitions, we can write the cross section in the FONLL scheme as
\beq\label{eq:sigma_matched}
\sigma^{\textrm{matched}} = \sigma^{(4)} + \sigma^{(5)} - \sigma^{(4-5)}\,.
\eeq
The fact that $\sigma^{(4-5)}$ removes the overlap between the cross sections computed in the 4FS and 5FS is guaranteed by noting that 
\beq
\label{eq:matching_consistency}
{B}^{(n)}_{ij}\!\left(z,L_f,L_r,m_b\right)
-{A}^{(n)}_{ij}\!\left(z,L_f,L_r,L_b\right) = \ord\left({m_b^2}\right)\,.
\eeq
Using a straightforward rearrangement of terms, we can cast eq.~\eqref{eq:sigma_matched} into the alternative form,
\beq\label{eq:sigma_matched_alternative}
\sigma^{\textrm{matched}} = \sigma^{(4)} + \tilde{\sigma}^{(5)} - \tilde{\sigma}^{(4-5)}\,,
\eeq
with 
\beq
\tilde{\sigma}^{(4-5)} = \tau\sum_{i,j=-4}^4\,\mathscr{L}_{ij}^{(5)}(\tau,\mu_F^2)\otimes \delta\eta^{(5)}_{ij}\!\left(\tau,L_f,L_r,L_b,\alpha_s^{(5)},y_b^2\right)\,,
\eeq
and $\tilde{\sigma}^{(5)}$ collects only those channels in the 5FS that have a $b$-quark in the initial state (we suppress again the dependence on all arguments for readability)
\beq\bsp
\tilde{\sigma}^{(5)} &\,= 2\,\tau\Bigg[\mathscr{L}_{b\bar{b}}^{(5)}\otimes {\eta}_{b\bar{b}}^{(5)} + \left(\mathscr{L}_{bg}^{(5)}+\mathscr{L}_{\bar{b}g}^{(5)}\right)\otimes {\eta}_{bg}^{(5)} +\frac{1}{2} \left(\mathscr{L}_{bb}^{(5)}+\mathscr{L}_{\bar{b}\bar{b}}^{(5)}\right)\otimes {\eta}_{bb}^{(5)}\\
&\,
+\sum_{q=1}^4 \left(\mathscr{L}_{bq}^{(5)}\otimes \eta_{bq}^{(5)}+\mathscr{L}_{\bar{b}q}^{(5)}\otimes \eta_{\bar{b}q}^{(5)}\right)+\sum_{\bar{q}=-4}^{-1} \left(\mathscr{L}_{b\bar{q}}^{(5)}\otimes \eta_{b\bar{q}}^{(5)}+\mathscr{L}_{\bar{b}\bar{q}}^{(5)}\otimes \eta_{\bar{b}\bar{q}}^{(5)}\right)\Bigg]\,.
\esp\eeq

With the completion of the N$^3$LO corrections in 5FS, we have now for the first time the possibility to compute all ingredients in eq.~\eqref{eq:sigma_matched} consistently through third order in the strong coupling. The phenomenological impact of these corrections will be explored in the remainder of this paper.

\section{Phenomenological results}
\label{sec:matched_pheno}

In this section we present our results for the inclusive cross section matched according to the FONLL procedure through third order in the strong coupling. We work with a Higgs mass of $m_H=125\textrm{ GeV}$ and the pole mass of the bottom quark is $m_b=4.58\textrm{ GeV}$. The strong coupling and the Yukawa coupling are evaluated at the renormalisation scale $\mu_R^2$ using three-loop running in the $\overline{\textrm{MS}}$-scheme~\cite{Tarasov:1980au,Larin:1993tp,vanRitbergen:1997va,Baikov:2016tgj,Herzog:2017ohr,Harlander:2003ai,Chetyrkin:1997dh,Czakon:2004bu,Baikov:2014qja}, and we start the evolution from $\alpha_s(m_Z^2)=0.118$ and $m_b(m_b)=4.18\textrm{ GeV}$. 
We choose to work with the PDF set of ref.~\cite{Bonvini:2016fgf,bonvinipdfs}, which is based on the combined \texttt{PDF4LHC15\_nnlo\_mc} set~\cite{Butterworth:2015oua}, but starting from a low scale where there is no bottom quark, and then performing the evolution to higher scales using a consistent value of the bottom pole mass throughout. 

The 4FS results are generated using {\sc\small MadGraph5\_aMC@NLO}~\cite{Alwall:2014hca}.
The computation of the one-loop amplitudes is carried out with the module {\sc\small MadLoop}~\cite{Hirschi:2011pa,Alwall:2014hca}, which generates the loop integrand using an in-house implementation of the {\sc\small OpenLoops} optimisation~\cite{Cascioli:2011va}.
The loop integrals are then evaluated by switching dynamically between two one-loop reduction techniques: OPP~\cite{Ossola:2006us} or Laurent-series expansions~\cite{Mastrolia:2012bu} that are performed at the integr\emph{and} level, and methods applied at the tensor integr\emph{al} level~\cite{Passarino:1978jh,Davydychev:1991va,Denner:2005nn}. These reduction techniques have been automated in tools that {\sc\small MadLoop} interfaces to: {\sc\small{CutTools}}~\cite{Ossola:2007ax}, {\sc\small{Ninja}}~\cite{Peraro:2014cba,Hirschi:2016mdz} and {\sc\small{COLLIER}}~\cite{Denner:2016kdg}. The renormalisation of the bottom quark Yukawa coupling is performed by default in the on-shell scheme in {\sc\small MadGraph5\_aMC@NLO}~\cite{Alwall:2014hca}. In order to renormalise this quantity in the $\overline{\textrm{MS}}$-scheme instead (and correctly account for the running of $y_b(\mu_R)$ in this case), we must perform adjustments\footnote{See \url{https://cp3.irmp.ucl.ac.be/projects/madgraph/wiki/bbH} for a comprehensive list of these changes.}
 of the process output identical to those considered in ref.~\cite{Wiesemann:2014ioa}.
Finally, we note that the top mass contributions of order $\mathcal{O}(y_b^2)$ (i.e. but \emph{not} the ones involving $y_t$)  are \emph{included} in the NLO 4FS computation (whereas they are \emph{not} in the N$^3$LO 5FS computation). These top-quark contributions come in through corrections of the triple-gluon vertex as well as the gluon propagator (and therefore its wavefunction counterterm).
We stress that considering the top-quark contribution only in the 4FS part of the computation does \emph{not} spoil the consistency of the matching procedure presented in section~\ref{sec:fonnll}. In addition, we have verified that its numerical impact is at the permille level only.

Before we present our results, let us briefly comment on different ways to implement the FONLL matching procedure. More specifically, in refs.~\cite{Forte:2015hba,Forte:2016sja} three different scenarios were considered:
\begin{itemize}
\item {\bf FONLL-A:}  All ingredients in eq.~\eqref{eq:sigma_matched} are included through $\ord(\alpha_s^2)$. This corresponds to matching the 5FS at NNLO to the 4FS at LO, and all collinear logarithms are resummed at next-to-next-to-leading logarithmic accuracy (NNLL). Phenomenological results for Higgs production in bottom-quark fusion using the FONLL-A prescription have first been obtained in ref.~\cite{Forte:2015hba}.
\item {\bf FONLL-B:}  The contributions from $\sigma^{(4)}$ and $\tilde{\sigma}^{(4-5)}$ in eq.~\eqref{eq:sigma_matched_alternative} are included through $\ord(\alpha_s^3)$, while $\tilde{\sigma}^{(5)}$ is included only through $\ord(\alpha_s^2)$. In this way the fixed-order NLO accuracy of the 4FS is retained, and all collinear logarithms are resummed at NNLL.
Phenomenological results using the FONLL-B prescription have first been obtained in ref.~\cite{Forte:2016sja}.
\item {\bf FONLL-C:}  All ingredients in eq.~\eqref{eq:sigma_matched} are included through $\ord(\alpha_s^3)$. This corresponds to matching the 5FS at N$^3$LO to the 4FS at NLO, so that all collinear logarithms are resummed at NNLL.
Phenomenological results using the FONLL-C are presented for the first time in this paper.
\end{itemize}

\begin{figure*}[!t]
\centering
\begin{subfigure}[t]{0.49\textwidth}
\includegraphics[width=\textwidth]{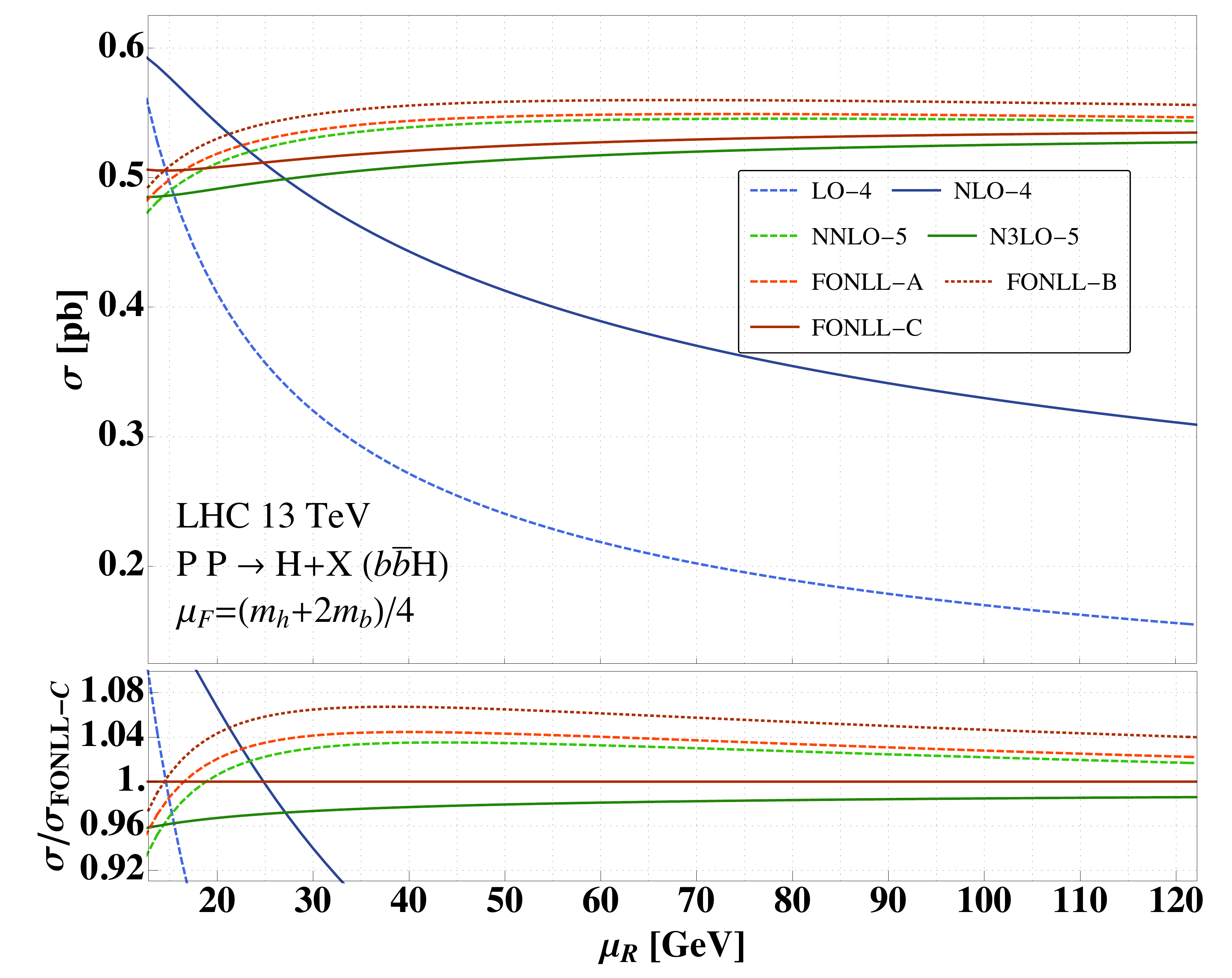}
\subcaption{
\label{fig:FONLMUR}
Variation of $\mu_R$.
}
\end{subfigure}
\begin{subfigure}[t]{0.49\textwidth}
\includegraphics[width=\textwidth]{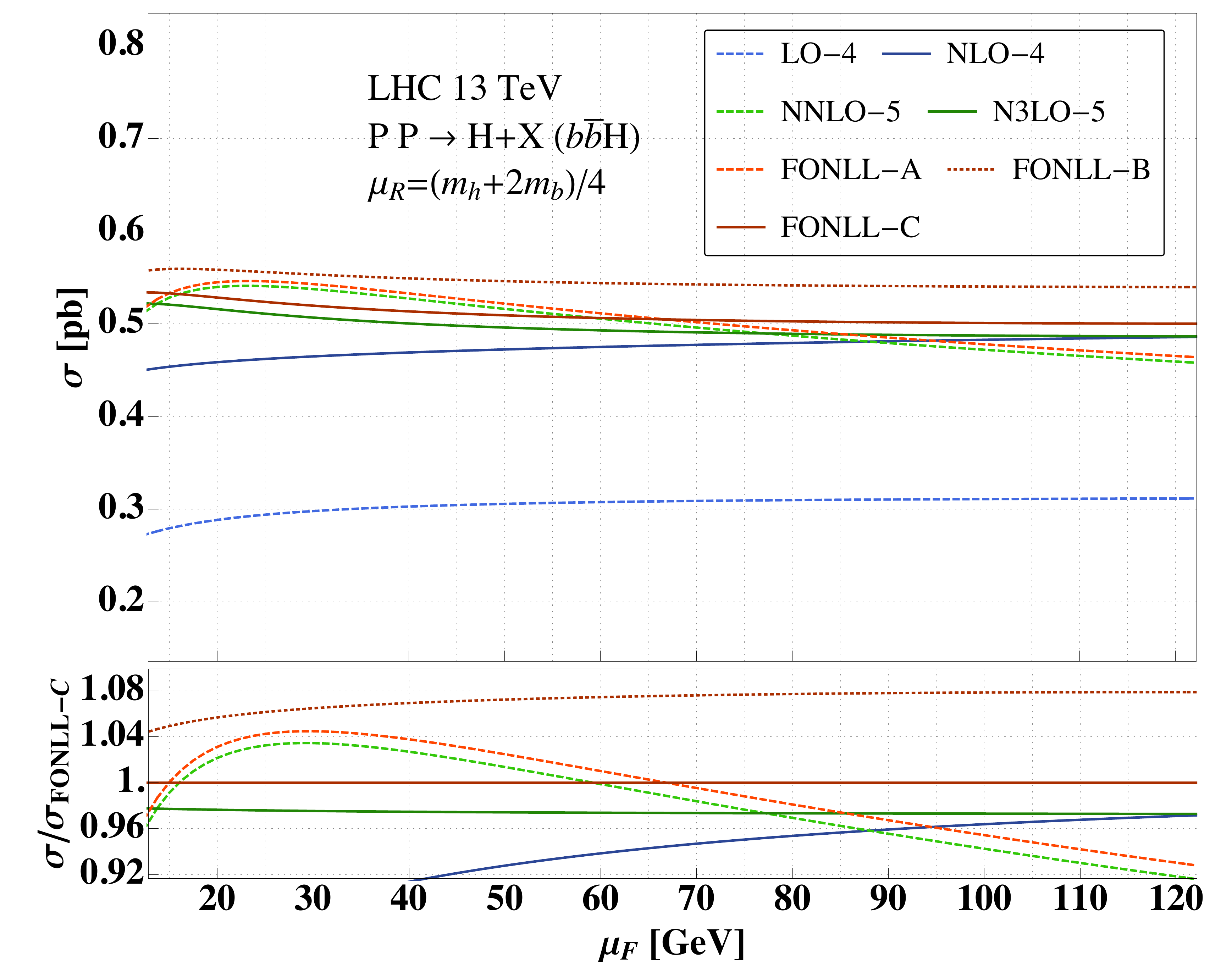}
\subcaption{
\label{fig:FONLMUF}
Variation of $\mu_F$.
}
\end{subfigure}
\caption{
Comparison of the four-, five-flavour and FONLL-matched cross sections. 
The fixed-order 4FS cross-sections (LO-4 and NLO-4) presented in these figures were obtained using the PDF set of ref.~\cite{Bonvini:2016fgf,bonvinipdfs} with the bottom mass set to infinity. In contrast, the 4FS cross-sections entering the FONLL matching procedure in eq.~\eqref{eq:4FSterms} were computed using the same PDF set as for the 5FS computation (i.e. PDF set evolved using a bottom mass set to 4.58 GeV).
}
\end{figure*}
In figs.~\ref{fig:FONLMUR} and~\ref{fig:FONLMUF} we show the variation of the 4FS, 5FS, and matched results with the renormalisation or factorisation scale, with the other scale held fixed. We observe that FONLL-C prediction increases the value of the N$^3$LO 5FS result by roughly 2\% over the whole range of scales considered, while maintaining the very reduced sensitivity to the residual scale dependence of the N$^3$LO result. This is at variance with the matching at the previous order (FONLL-A), where the matched prediction only resulted in a tiny increase of the 5FS cross section at NNLO~\cite{Forte:2015hba}. Finally, we observe that the FONLL-B prescription leads to a substantial increase of the cross section compared to the 5FS NNLO result. The FONLL-B prescription misses the contributions the $b$-initiated channels at N$^3$LO, which give large and negative contributions to the cross section. 
More precisely, the FONLL-B prescription does no satisfy eq.~\eqref{eq:matching_consistency} since it considers all 4FS contribution $B_{ij}^{(3)}$ while ignoring (i.e. effectively setting to zero) the 5FS counterpart pieces $\eta_{bi}^{(5,3)}$, $\eta_{\bar{b}i}^{(5,3)}$, $\eta_{ib}^{(5,3)}$ and $\eta_{i\bar{b}}^{(5,3)}$ contributing to $A_{ij}^{(3)}$.
As a consequence, it seems that for this particular process the FONLL-B prescription does not give a reliable estimate of the value of the cross section at $\ord(\alpha_s^3)$. This underlines the need to include the N$^3$LO 5FS prediction.

\begin{table}[h!]
\centering
\begin{tabular}{c | c | c | c | c | c  }
\hline\hline
$S$ [TeV] & $\sigma$ [pb] & $\delta(\text{scale})$ [\%] & $\delta(\alpha_S+\text{PDF})$ [\%]  & $\delta(\text{PDF-TH})$  [\%] & $\delta(m_b)$  [\%] \\
\hline
7 & 0.172 & ${}^{+2.50}_{-2.63}$ & $\pm$ 9.05 & $\pm$3.85 & ${}^{+1.44}_{-0.95}$ \\ 
\hline 
8 & 0.222 & ${}^{+2.64}_{-3.01}$ & $\pm$ 9.02 & $\pm$3.54 & ${}^{+1.44}_{-0.95}$ \\ 
\hline 
13 & 0.535 & ${}^{+2.52}_{-4.11}$ & $\pm$ 8.37 & $\pm$2.49 & ${}^{+1.44}_{-0.95}$ \\ 
\hline 
14 & 0.604 & ${}^{+2.67}_{-4.31}$ & $\pm$ 8.31 & $\pm$2.36 & ${}^{+1.44}_{-0.95}$ \\ 
\hline 
27 & 1.68 & ${}^{+2.57}_{-5.92}$ & $\pm$ 7.59 & $\pm$1.22 & ${}^{+1.44}_{-0.95}$ \\ 
\hline 
100 & 9.21 & ${}^{+3.26}_{-9.38}$ & $\pm$ 6.68 & $\pm$1.00 & ${}^{+1.44}_{-0.95}$ \\ 
\hline \hline
\end{tabular}
\caption{
\label{tab:numbers}
FONLL-C (N$^3$LO 5FS matched to NLO 4FS) predictions for the $b\bar b H$ cross section at different collider energies and associated uncertainties.
}
\end{table}
In tab.~\ref{tab:numbers} we present results for the matched cross section for various representative collider energies. 
We estimate the uncertainty due to the truncation of the perturbative series by varying the factorisation and renormalisation scales independently up and down by a factor around the central values $(\mu_F,\mu_R) = ((m_H+2m_b)/4,m_H)$ within the constraint of eq.~\eqref{eq:scaleconstraint}. 
This choice for the central scales was discussed in section~\ref{sec:5flav}. 
Furthermore, we quote the PDF and strong coupling uncertainty $\delta(\alpha_S+\text{PDF})$, the PDF theory uncertainty $\delta(\text{PDF-TH})$ and the bottom quark mass uncertainty $\delta(m_b)$ that we asses based on the five-flavour cross section as outlined in section~\ref{sec:5flav}.

%%%%%%%%

%\input{ThresholdExpansion.tex}
%\input{Renormalisation.tex}
%\input{IR.tex}

% !TEX root = paper.tex

\section{Conclusion}
\label{sec:conclusion}

In this paper we have performed a detailed phenomenological study of Higgs production in bottom quark fusion.
In a first part of the paper we have focused on the N$^3$LO cross section in the 5FS.
We described the structure of the analytic partonic coefficient function for this cross section as well as for the matching contribution $\tilde{\sigma}^{(4-5)}$ and include it in electronically readable form together with the arXiv submission of this article.
Next, we elaborated on the phenomenological analysis of ref.~\cite{Duhr:2019kwi}. 
We have studied the dependence of the cross section of the renormalisation and factorisation scales. 
We observe a convergent behaviour of the perturbative series, provided that the factorisation scale is set to a relatively low value. 
This corroborates similar conclusions drawn based on the behaviour of the cross section at lower orders, and gives further support for this unconventionally low choice of the factorisation scale. 
We have also studied other sources of uncertainty that may affect our prediction for the cross section, including the effects due to PDFs and the strong coupling constant, as well as the value of the bottom quark mass that is used in the computation.

In a second part of the paper we have combined our N$^3$LO computation in the 5FS with the NLO cross section in the 4FS computed with {\sc\small MadGraph5\_aMC@NLO}. 
The overlap between the two schemes is removed using the FONLL matching procedure, first applied to Higgs production in bottom quark fusion in refs.~\cite{Forte:2015hba,Forte:2016sja}. 
The novelty of our computation lies in the fact that for the first time we can compute all quantities that enter the combination consistently through third order in the strong coupling. 
We find that the effect of the matching is non-negligible, increasing the value of the 5FS N$^3$LO cross section by roughly 2\%. 
We note that this increase is of the same order as the scale dependence at N$^3$LO. 
We also find that previous attempts to match the two schemes through third order in the strong coupling without including the complete N$^3$LO calculation had led to a substantially different answer. 
The reason is that the $b$-initiated channels at N$^3$LO give a large and negative contribution to the cross section, an effect which was not captured by previous calculations.

To conclude, we have presented the most precise prediction for the inclusive bottom quark fusion cross section by combining the most precise calculations in both the 4FS and 5FS. 
The non-negligible effect of the N$^3$LO corrections underlines once more the need for calculations at this order for the precision physics program at the LHC, and we expect that our results will play a role in the study of the interactions of the bottom quark and the Higgs bosons, both at the LHC and at future hadron colliders. 

\section*{Acknowledgements}
We would like to thank Stefano Forte and Davide Napoletano for useful discussion.
VH acknowledges the hospitality of the Theory Department at CERN at the early stages of this project.
This work is supported in part by the European Research Council grants No 637019 ``MathAm'' (CD) and No 694712 ``PertQCD'' (VH).
BM is supported by the Pappalardo fellowship.
FD was supported by the Department of Energy, Contract DE-AC02-76SF00515.

\bibliography{Bib,BibVH}
\bibliographystyle{JHEP}
\end{document}